\documentclass[a4paper,fleqn,usenatbib]{mnras}

\usepackage[T1]{fontenc}
\usepackage{ae,aecompl}

\usepackage{graphicx,color}	
\usepackage{amsmath}	
\usepackage{amssymb}	

\newcommand\mj{{\,{\rm M}_{\rm J}}}
\newcommand\msun{{\,{\rm M}_{\odot}}}

\title[Gas accretion onto planets]{A desert of gas giant planets beyond tens of au}

\author[Sergei Nayakshin]{
Sergei Nayakshin
\\
Department of Physics and Astronomy, University of  Leicester, Leicester LE1 7RH, UK. {E-mail: sn85@le.ac.uk}
}

\date{Accepted XXX. Received YYY; in original form ZZZ}

\pubyear{2015}

\begin{document}
\label{firstpage}
\pagerange{\pageref{firstpage}--\pageref{lastpage}}
\maketitle

\begin{abstract}
 Direct imaging observations constrain the fraction of stars orbited by gas giant planets with separations greater than 10 au to about 1\% only. This is widely believed to indicate that massive protoplanetary discs rarely fragment on planetary mass objects. I use numerical simulations of  gas clumps embedded in massive gas discs to show that these observations are consistent with $\sim 0.2 - 10$ planetary mass clumps per star being born in young gravitationally unstable discs.  A trio of processes -- rapid clump migration, tidal disruption and runaway gas accretion -- destroys or transforms all of the simulated clumps into other objects, resulting in a desert of gas giants beyond separation of approximately 10 au. The cooling rate of the disc controls which of the three processes is dominant. For cooling rates faster than a few local dynamical times, clumps always grow rapidly and become massive brown dwarfs or low mass stars. For longer cooling times, post-collapse (high density) planets migrate inward to $\sim 10-20$~au where they open a gap in the disc and then continue to migrate inward much less rapidly. Pre-collapse (low density) planets are tidally disrupted and may leave massive solid cores behind. Gas giant planets observed inside the desert, such as those in HR 8799, must have followed an unusual evolutionary path, e.g., their host disc being dispersed in a catastrophic fashion. 
\end{abstract}


\section{Introduction}

Evolution of protoplanetary discs is an active area of research \citep[for recent reviews see][]{KratterL16,HaworthEtal16}. One of the open questions is the self-gravitational instability \citep{Toomre64} of protoplanetary discs, which may \citep{Kuiper51b,Boss97} or may not \citep[e.g.,][]{Kumar72,GoldreichWard73,LaughlinBodenheimer94} lead to formation of gas giant planets by fragmentation of the discs on clumps of a few Jupiter mass \citep[see][for a review]{HelledEtal13a}.

\cite{Gammie01} has shown that self-gravitating gas discs fragment only when the disc cooling time, $t_{\rm cool}$, is shorter than a few local dynamical times, e.g., $t_{\rm cool} \lesssim \beta/\Omega_{\rm k}$, where $\Omega_{\rm k}$ is the local Keplerian frequency and $\beta \sim 3-6$ \citep[see also][]{Johnson03,Rice05,CossinsEtal09}. \cite{Rafikov05} used this fact to constrain the location of disc fragmentation to separations $R \gtrsim 50$~au \citep[see also][]{DurisenEtal07,Boley09,RogersWadsley12}.

More recent simulations suggest that gas discs may fragment at higher values of $\beta$ \citep[e.g.,][]{MeruBate11a,MeruBate12,Paardekooper12a}, and it is currently not clear whether there is a well defined value of critical $\beta$ \citep{MichaelEtal12,RiceEtal12,RiceEtal14a} or the fragmentation is stochastic in nature \citep{YoungClarke16,LinKratter16}. Nevertheless, there is a general agreement that discs are not likely to fragment closer than tens of au to the host star \citep[see][]{KratterL16}.


The outcome of gravitationally unstable disc fragmentation is even more controversial. Most authors \citep[e,g.,][]{SW08,KratterEtal10,ForganRice11,ForganRice13,TsukamotoEtal14} find that the initial fragment mass is typically too high to be relevant to formation of gas giants that are observed to be dominated by planets with masses $\sim 1\mj$ \citep[e.g.,][]{MayorEtal11}, leading to formation of brown dwarfs rather than planets. This may however be debatable since the current uncertainty in the exact disc fragmentation conditions translates into a much larger uncertainty in the initial fragment mass \citep[\S 4.3 in][]{Nayakshin_Review}; some authors do find $\sim 1 \mj$ fragments forming in their discs \citep[e.g.,][]{BoleyEtal10}.

Evolution of gas clumps after their formation is also highly uncertain. Gas clumps of planetary mass are found to migrate inward on time scales of just a few thousand years, depending on the disc and planet masses and other parameters \citep{MayerEtal04,VB06,BoleyEtal10,InutsukaEtal09,MachidaEtal11,BaruteauEtal11,MichaelEtal11,ChaNayakshin11a}. Further, the clumps may evolve not only in separation but also in mass. If they accrete gas rapidly \citep{KratterEtal10,ZhuEtal12a,ForganRice13b}, they become brown dwarfs or even low mass stellar companions. The reverse is possible too. If clumps do not accrete gas and contract slowly, then they are tidally disrupted \citep{BoleyEtal10,Nayakshin10c} after they migrate too close to the host star, leaving behind a solid core \citep[if grain sedimentation within the clump was sufficiently rapid; see][]{Boss97,HS08,HelledEtal08,Nayakshin10b}. This channel of clump evolution is the base for the Tidal Downsizing scenario of planet formation \citep[see the recent review in][]{Nayakshin_Review}.

In this paper, the role of the cooling rate of the gas inside the Hill sphere of the clump in determining its evolutionary path is investigated. First, a simple analytical argument is made. Accretion of gas onto the embedded clumps must be inefficient when the gas cooling rate is long compared with $\Omega_{\rm k}^{-1}$. The validity of the argument is then investigated numerically, and confirmed, in the setting of the "$\beta$-cooling" self-regulating discs.

The $\beta$-cooling model is useful in its clarity and simplicity of comparison to analytical arguments, and has been used by a great number of authors, but real astrophysical flows have cooling times depending on the local gas density and temperature rather than gas distance from the host star. Therefore, in the next part of the paper I consider a protoplanetary disc with an approximate but physically better motivated cooling model. The long and the short cooling time regimes are explored in this model by varying a dust opacity multiplier, meant to capture the significant microphysical uncertainties in dust grain physics \citep[e.g.,][]{SemenovEtal03,DD05,HB11}. In addition, both pre-collapse (low density) and post-collapse (high density) gas giant planets are considered (see \S \ref{sec:num_planet} for detail). A wide range of initial gas clump masses, from $0.5 \mj$ to $16 \mj$, is covered.

Although the results of these more realistic runs vary significantly with parameters and physics of the simulations, one robust result stands out. A gas clump of a few Jupiter mass embedded in a massive gas disc at $\sim 100$~au evolves away from that position in either separation or mass, or both, only within some 10,000 years. This is much shorter than the typical $\sim 3$ Million year disc dispersal time of protoplanetary discs \citep{HaischEtal01}, implying that self-gravitating discs are {\it extremely unlikely} to leave giant planets beyond tens of au after the disc is dissipated away. This implies that the directly imaged gas giant planets must represent only a tip of the iceberg (perhaps as small as $\sim 0.1$\%) of a much larger population of gas clumps that existed there during the planet formation epoch.



\section{Analytical Arguments}\label{sec:analytics}

Let the planet be on a circular Keplerian orbit around the star a distance $R_0$ away from it. In the non-inertial system of coordinates rotating together with the planet, the relative azimuthal velocity of a circular Keplerian flow at radius $R_0 + \Delta R$ is $\Delta v = - (3/2) \Omega_K \Delta R$. This implies that the time scale for the differential gas flow to cross the Hill's sphere of the planet is 
\begin{equation}
t_{\rm cross} = {R_{\rm H} \over \Delta v(R_{\rm H})} \sim {1\over \Omega_K}\;.
\label{tc1}
\end{equation}
The gas enters the planet Hill sphere with a positive total energy. To become gravitationally bound to the planet, this energy needs to be reduced by radiative cooling and become negative before the gas leaves the Hill sphere on the other end. Defining a cooling time scale for the gas, $t_{\rm cool}$, as
\begin{equation}
t_{\rm cool} \equiv { u \over \Gamma}\;,
\end{equation}
where $u$ is gas specific energy and $\Gamma$ is the specific energy loss rate by radiation, we should expect that gas accretion onto the planet will be efficient if 
\begin{equation}
t_{\rm cool} \ll t_{\rm cross}\;,
\label{tc2}
\end{equation}
and inefficient in the opposite case, $t_{\rm cool} \gg t_{\rm cross}$. In the latter limit the gas enters the Hill sphere only temporarily. After entering the sphere, it is compressed and heats up nearly adiabatically. It is very likely to leave the Hill sphere since it is too hot to be bound to the planet.

\section{Massive self-gravitating $\beta$ discs}\label{sec:beta}

\subsection{Numerical setup}

\subsubsection{Gas discs}

\cite{BaruteauEtal11} studied migration of massive gas giant planets in self-gravitating discs in which radiative cooling of the disc is described by the widely used "$\beta$-cooling" model \citep[e.g.,][]{Gammie01,Rice05}:
\begin{equation}
{du \over dt} = - {u\over t_{\rm cool}}\;,
\label{beta0}
\end{equation}
where the cooling time $t_{\rm cool}$ is a function of radius $R$ only, described by
\begin{equation}
t_{\rm cool}(R) = \beta \Omega_K^{-1}\;,
\label{beta_def}
\end{equation}
where $\beta$ is a positive constant and $\Omega_K = (GM_*/R^3)^{1/2}$ is the Keplerian angular frequency at radius $R$. They considered $\beta=10$, 15, and 30, for which their discs are in a self-regulating but not fragmenting gravito-turbulent state \citep{Gammie01}.

I follow a similar setup here with an added focus on capturing gas accretion onto the planets. The disc has an initial disc surface density profile $\Sigma(R) \propto R^{-3/2}$ defined in the radial range $ 20 \le R \le 200$~AU, with the total gas mass $M_{\rm d} = 0.4\msun$. The disc is orbiting the central star of mass $M_* = 1\msun$, which is modelled as a sink particle with accretion radius $r_{\star} = 5$~AU. Initially, no planet is present in the disc, and the disc is relaxed for tens of orbits on the outer edge, so that it settles into the self-regulating quasi-steady state \citep{Gammie01}. During this relaxation, some of the disc material is accreted onto the star, so that the disc mass decreases to $M_{\rm d} \approx 0.35$ (typically).

\subsubsection{Treatment of the planet}\label{sec:num_planet}

Radiative contraction of a gas giant planet formed by gravitational disc instability is conveniently divided into the pre-collapse and the post-collapse phases \citep{Bodenheimer74}. The former is relevant to the gas clumps that have only recently formed out of the surrounding protoplanetary disc. They are dominated by molecular Hydrogen, spatially extended and thus have low gas densities. They are therefore susceptible to the tidal disruption by the host star at separations as large as a few au to tens of au \citep{BoleyEtal10}. 

After the clumps get rid of their excess formation heat and their central temperature reaches $\sim 2000$~K, H$_2$ molecules dissociate, and the planet collapses dynamically to much higher densities. This process is similar to the second collapse in star formation \citep{Larson69}. The post-collapse clumps (planets) are orders of magnitude denser and hotter. Such planets may be tidally disrupted only at $\sim 0.1$~au separations from the star, which we do not study here.

For the $\beta$-cooling discs, only pre-collapse planets are studied. The clumps are modelled explicitly with SPH rather than with a sink particle approach \citep[e.g.,][]{Bate95}. The latter is best suited for problems in which gas is accreting onto the sink at nearly the free-fall velocity, such as the classical Bondi-Hoyle accretion problem; for accurate results the sink radius, $r_{\rm sk}$, must be picked much smaller than the Bondi capture radius \citep[see][]{Cuadra06}. Radius of pre-collapse planets is typically only a factor of a few times smaller than the Hills radius. Gas captured by the planet from the surrounding disc settles onto it at sub-sonic velocities rather than at free-fall, as we shall see later on. For both of these reasons the sink particle approach is not desirable.

The evolution of pre-collapse planet internal structure depends on many physical variables and physical effects \citep[e.g.,][]{HB11,VazanHelled12,Nayakshin15a,Nayakshin16a} that cannot be yet simulated self-consistently in 3D. A very simple model is therefore employed to model the clump and its contraction. To initialise the planet, a uniform density gas sphere is inserted in the gaseous disc that was relaxed as described in the previous section. The mass of the sphere is $M_{\rm p0} = 3\mj$, and its radial extent is $r_0 = 3$~AU. The initial clump temperature is set to 10 K. With these initial conditions, the clump is strongly gravitationally bound and initially collapses dynamically to higher density. The collapse slows when adiabatic compression heats the gas up so that the pressure gradient is sufficiently large to oppose clump self gravity. Further contraction of the clump is governed by the rate at which it looses its thermal energy content. 

Detailed stellar evolution calculations show that the contraction time of pre-collapse Jovian planets is on the order of $\sim 10^5-10^6$ years \citep[e.g.,][]{Bodenheimer74,HelledEtal08}. The simple $\beta$-cooling law (equations \ref{beta0} and \ref{beta_def}) is clearly not appropriate inside the planet. Therefore, to avoid an unphysically fast contraction of the planet, the cooling law is modified at high densities,
\begin{equation}
t_{\rm cool} = \beta \Omega_K^{-1} \left(1 + {\rho \over \rho_{\rm crit}}\right)\;,
\label{beta_mod}
\end{equation}
where $\rho_{\rm crit}$ is set to $10^{-10}$ g~cm$^{-3}$ for most tests in this section. This value of the density is $\sim 1000$ times higher than the tidal density (that is, the maximum density of a $Q=1$ disc) of the disc at $R=100$~AU, so that the modification is only significant at the surface or inside the planet.

The clump is inserted into the disc on a local circular orbit in the disc midplane at separation $R_0 = 100$~AU from the star. For definitiveness, the current mass of the planet is defined in the simulations  as the total mass of SPH particles within distance $R_{\rm c}= 0.5 R_{\rm H}$ from the clump centre. This definition is chosen empirically, based on the result that the material found within that region is usually strongly bound to the planet and does not becomes unbound, whereas gas farther away from the planet centre is much more likely to be unbound. Changes in the definition of $R_{\rm c}$ to either smaller or larger values do not affect the main conclusions of this paper quantitatively.

\subsection{The fiducial $\beta=7.5$ case}\label{sec:beta7.5}

I begin by studying the case with $\beta = 7.5$. The initial SPH particle number is $N_{\rm sph} = 400,000$ (before the disc relaxation). The SPH particle mass is hence $m_{\rm sph} = 10^{-6} \msun$. The planet is inserted at the location $(x,y,z) = (100, 0, 0)$~au at time $t=0$. 

\begin{figure*}
\includegraphics[width=0.99\columnwidth]{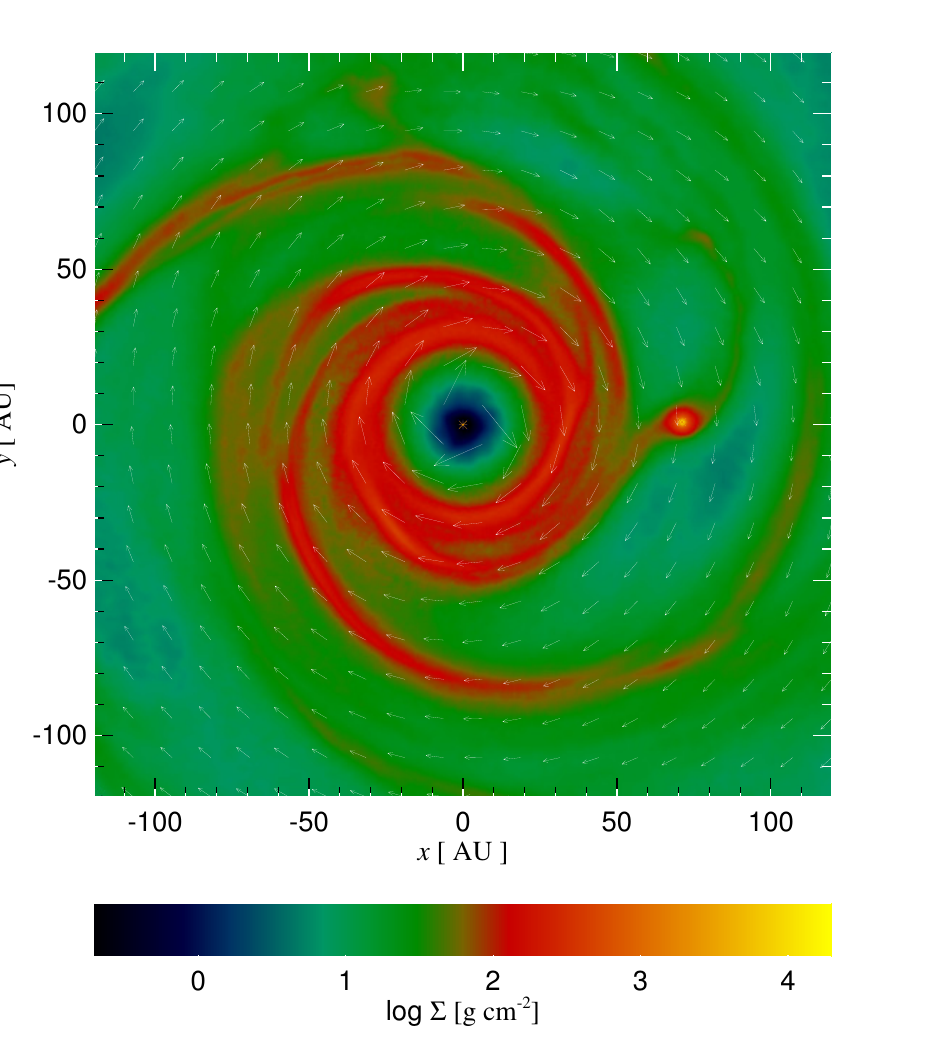}
\includegraphics[width=0.99\columnwidth]{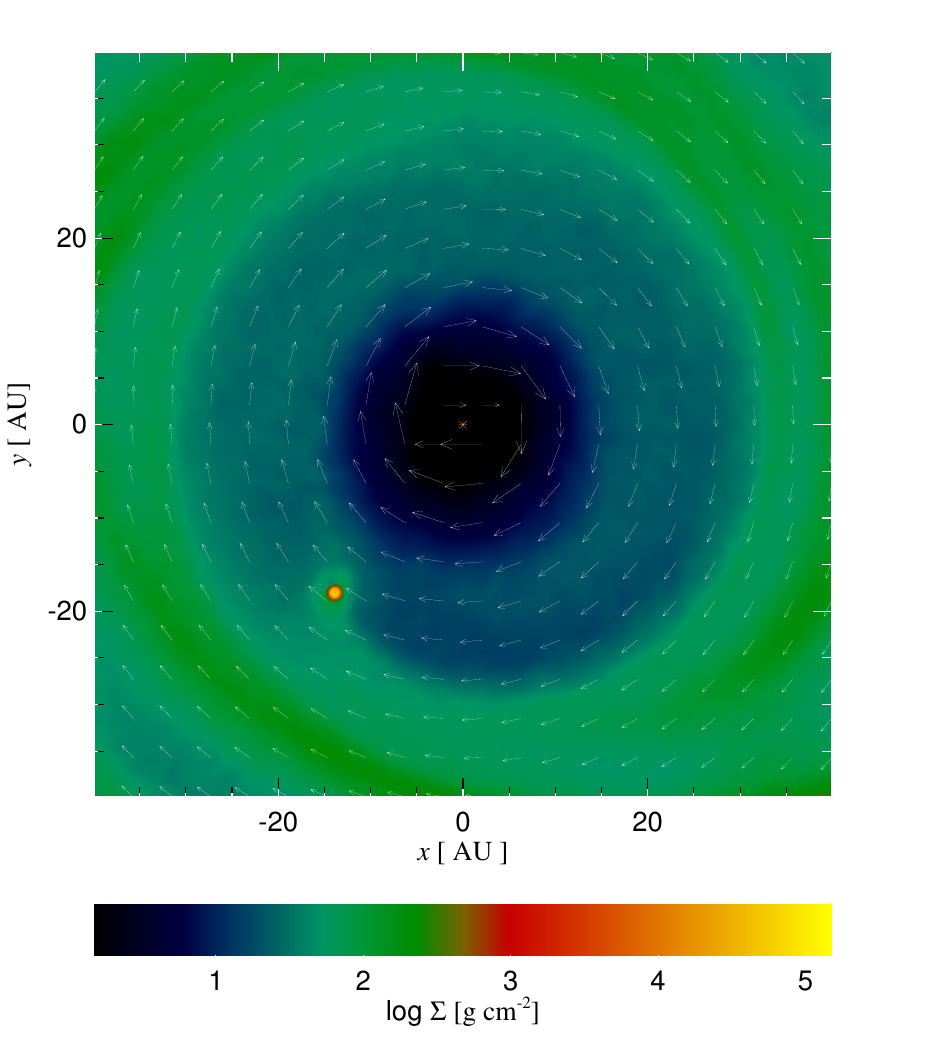}
 \caption{Face on projection of the gas surface in the fiducial simulation $\beta7.5$ at time $t=670$ (left panel) and $t=5600$ (right panels) years. The planet is the highest surface density peak, and migrates inward to the inner edge of  the disc within $\sim 1500$ years. The white arrows show the gas velocity field. The asterisk in the centre of the panels is the host star.}
   \label{fig:beta7.5_snap}
 \end{figure*}

Figure \ref{fig:beta7.5_snap} shows the map of the column density of the disc with the embedded planet after the latter made one revolution around the star, which corresponds to time 670 years, in the left panel. The right panel shows the same at much later time, $t = 5600$~years. The star is located in the centre of both panels. The overlying white vector field shows the gas velocity flow map, normalised on the largest velocity found in the figure.

The clump migration rate is very high in this simulation (the separation and the mass  versus time tracks of the planets are shown in fig. \ref{fig:RES}). The planet has already migrated to radius $R= 71.4$~AU in the left panel of Fig. \ref{fig:beta7.5_snap}, after just one rotation. In this particular simulation, the planet migrates close to the inner boundary of the disc (20 AU) in some $\sim 1500$~years, but then practically stalls there because there is little gas material there to interact with. This stalling of the planet is probably mainly due to the initial conditions chosen -- the planet is likely to continue to migrate inward had the disc been present at smaller radii. On the other hand, 1D disc-planet migration calculations show that gas giant planets usually open deep gaps in the disc when they reach the radius of $\sim 10-20$~AU \citep[e.g., see figs. 2 \& 3 in][]{Nayakshin15d}, and then continue to migrate slower in the type II regime. Therefore we should expect the planet migration to slow down after reaching these radii in a full disc as well, although perhaps at slightly smaller radii. Understanding the eventual fate of the planet requires simulations on much longer time scales of disc removal, $\sim 3 $ Million years, which is not achievable due to numerical limitations in 3D simulations.

\begin{figure}
\includegraphics[width=0.99\columnwidth]{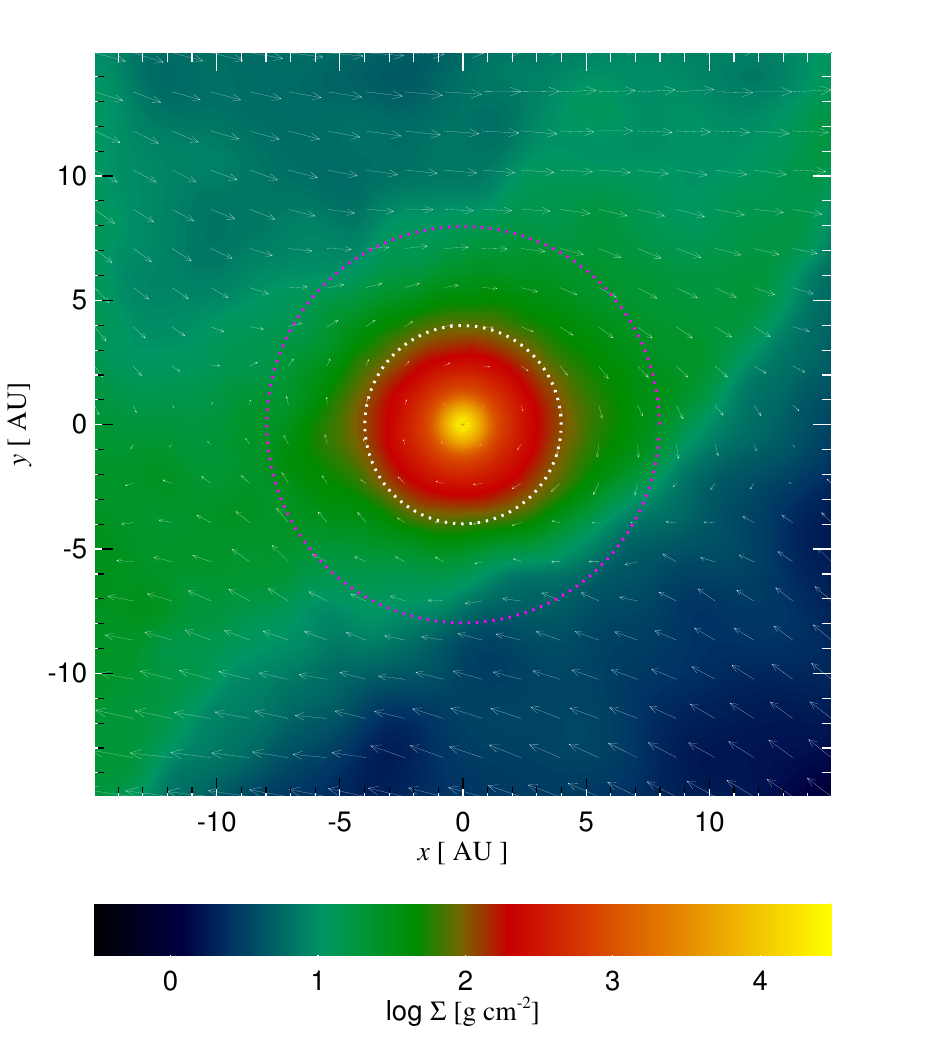}
 \caption{Zoom in on the region close to the planet at time $t=670$ years. The circles drawn in the figure are centred onto the densest part of the planet and have radii of $R_{\rm H}/2$ and $R_{\rm H}$. For detail see text in \S \ref{sec:beta7.5}.}
   \label{fig:beta7.5_zoom}
 \end{figure}

Figure \ref{fig:beta7.5_zoom} shows a zoom-in view of the region around the planet at time $t=670$~years, with the dotted curves showing the circles of radii $R_{\rm H}$ and $0.5 R_{\rm H}$ (magenta and while colours, respectively) centred on the densest part of the planet. For the latter plot, only SPH particles within a slab $|z| < 3$ AU are used. The gas mass within the white circle is $4.4\mj$, which is almost 50\% larger than the planet initial mass. Thus, the planet not only migrated but also increased in mass significantly in just one revolution around the star. 

The velocity vectors, while showing prograde circulating motions around the centre of the planet, decrease rather than increase towards the centre of the planet. The velocity field is hence not Keplerian. This is because the material around the planet's location is supported against planet self-gravity mainly by the gas pressure gradient rather than by rotation.

\subsubsection{Numerical resolution}\label{sec:res}

\begin{figure}
\includegraphics[width=0.99\columnwidth]{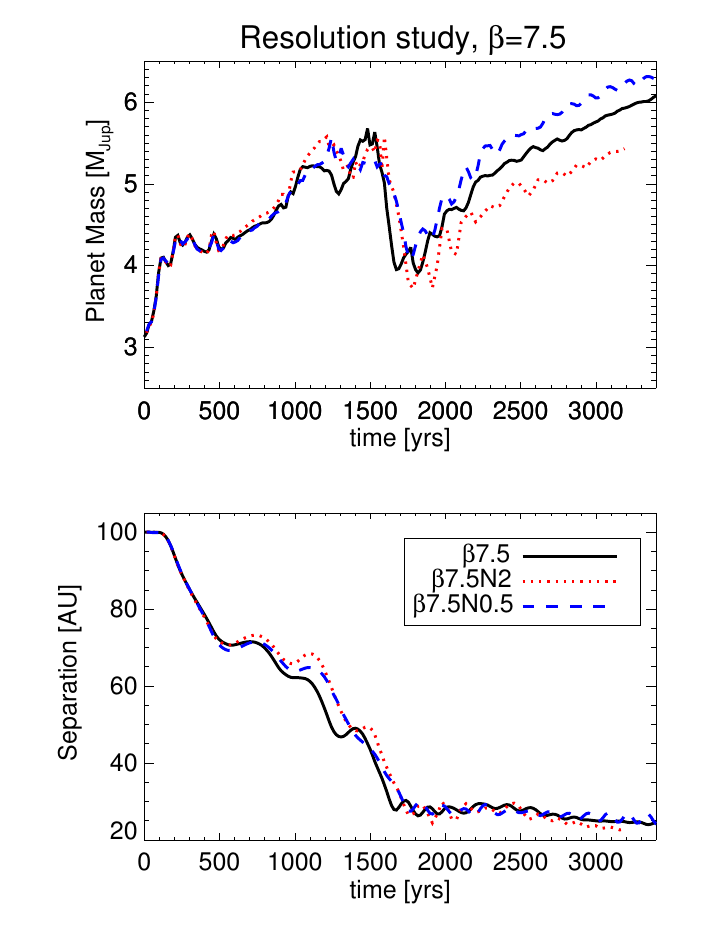}
 \caption{Planet mass (top panel) and separation to the host star (bottom panel) versus time in a resolution study. The previously studied case, $\beta=7.5$, with SPH particle mass $m_{\rm sph} =10^{-6} \msun$, shown with the solid black curves, is compared with two simulations differing from it only by the number of particles. There are twice as many and twice as few SPH particles for the  dotted and the dashed curves, respectively.}
   \label{fig:RES}
 \end{figure}

Fig. \ref{fig:RES} shows the results of the already presented fiducial simulation, $\beta7.5$, and two other simulations that are different only by the number of the SPH particles used. The dotted red curves are for simulations with twice as many SPH particles whereas the dashed blue ones are for twice as few. The top and the bottom panels of the figure show the planet mass and the planet-star separation, respectively, versus time.

We see that planet separation shrinks from 100 au to about 25 au in just 1500 years in all three of the simulations. There are some differences between the three sets of curves, but the variation is at the level of  $\sim$10\% between the three simulations. The planet mass changes are slightly larger than that. On the whole, while these differences are not negligible, they are acceptably small because it will be seen later on that both accretion and migration rates of the planet vary much more significantly when the physical parameters of the problem, such as $\beta$, $\rho_{\rm crit}$, planet's initial radius or density, and even the starting azimuthal angle,  are varied. This suggests that numerical resolution of the fiducial run ($N_{\rm sph} = 400,000$) is sufficient for our purposes here.


 \subsubsection{Dependence on the planet size}\label{sec:size}
 
 Another source of uncertainties in the outcome of these simulations is the internal physics of the planet, e.g., the planet physical size and the planet contraction rate. To investigate these issues, the initial conditions for the perturbation inserted into the disc, and the critical density at which the radiative cooling of gas is suppressed, $\rho_{\rm crit}$, were varied. In simulation $\beta$7.5T0.5, the initial temperature of the perturbation at time $t=0$ was halved (to $T = 5$~K). In simulation $\beta$7.5R2, the same change is made, and the initial perturbation radius also doubled, to $r_0 = 6$~AU. To test how the contraction rate of the planet affects the outcome, simulation $\beta$7.5C9 is started from the same initial condition as simulation $\beta$7.5, but the cooling time prescription (eq. \ref{beta_mod}) uses $\rho_{\rm crit} = 10^{-9}$ g~cm$^{-3}$ instead of $10^{-10}$ g~cm$^{-3}$. This allows the planet to contract more rapidly.
 
\begin{figure}
\includegraphics[width=0.99\columnwidth]{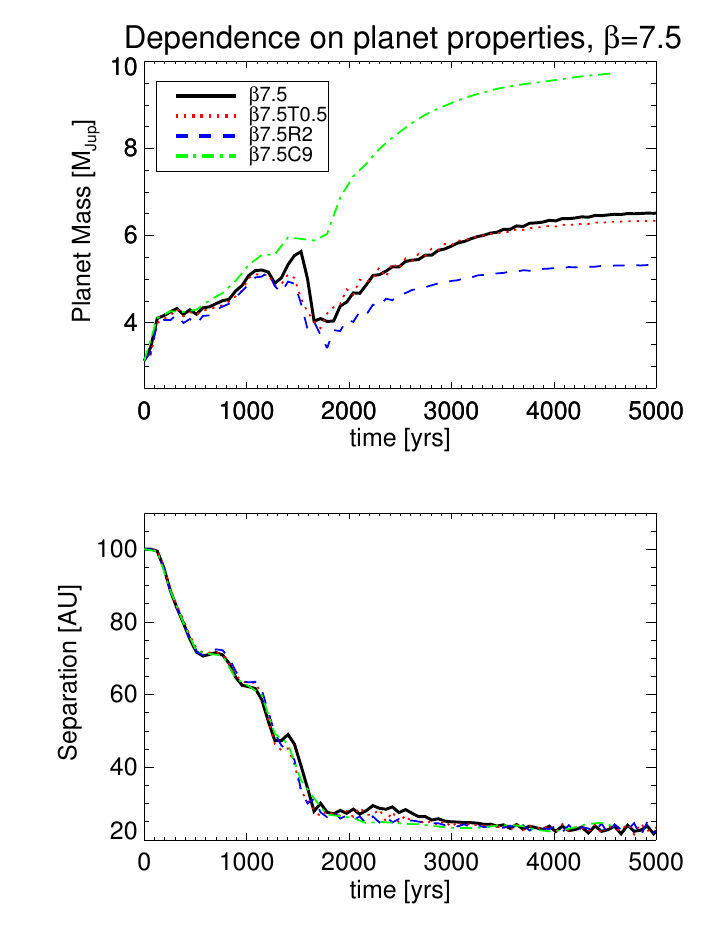}
 \caption{Dependence of results on the internal properties of the planet. See \S \ref{sec:size} for detail.}
   \label{fig:SIZE}
 \end{figure}

Fig. \ref{fig:SIZE} shows how these three simulations differ from the standard simulation ($\beta$7.5, reproduced with the black curve for comparison). The results of simulation $\beta$7.5T0.5 are hardly different from that of $\beta$7.5. This is so because the planet density and radius right after the initial collapse of the clump at $t\approx 0$ do not vary significantly between these two runs. Simulation $\beta$7.5R2, on the other hand, yields a less dense clump since the initial density of the perturbation is much lower. This implies that at a given mass, a more extended clump accretes material at a lower rate. 

Thus, with other things being equal, a more compact planet should accrete gas at a higher rate\footnote{A cautionary note here is that these simulations do not include the radiative feedback from the protoplanet, intensity of which should in general be larger for more compact planets. As was shown by \cite{NayakshinCha13} and \cite{Stamatellos15}, radiative feedback from the planet slows down accretion of gas onto it by heating up the surrounding gas and making it less bound to the planet.}. This is born out by simulation $\beta$7.5C9 in which the planet manages to accrete far more gas than it does in the fiducial simulation. The most likely physical reason for this result is simply the fact that a more compact planet allows a denser atmosphere to be built around the planet within the Hill sphere. This atmosphere is bound to the planet stronger, encouraging more gas to get accreted.

\begin{figure}
\includegraphics[width=0.99\columnwidth]{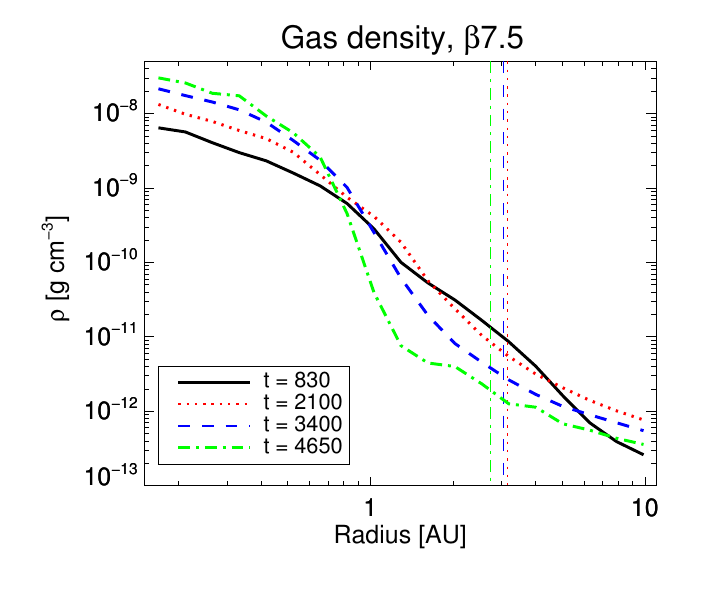}
\includegraphics[width=0.99\columnwidth]{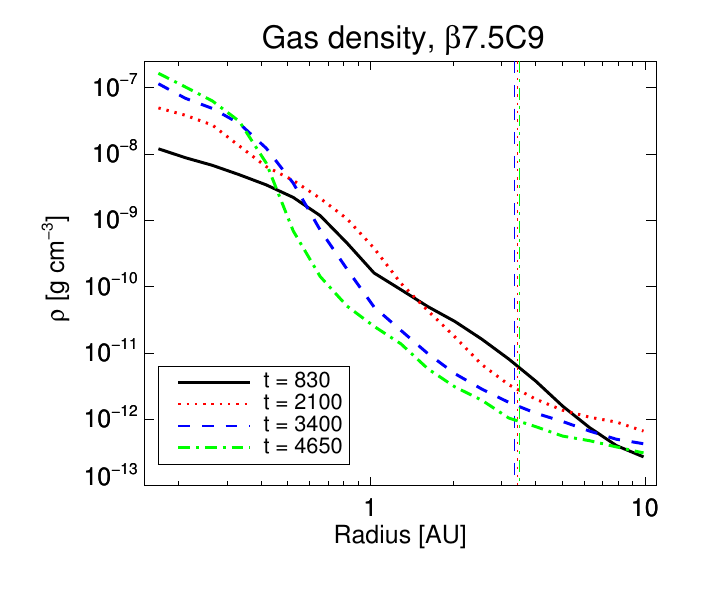}
 \caption{Gas density as a function of distance from the planet centre for the fiducial simulation $\beta$7.5 (top panel), and the one with faster internal planet cooling, $\beta$7.5C9 (bottom panel). The latter contracts faster, as expected (see \S \ref{sec:size}).}
   \label{fig:beta7.5_planet}
 \end{figure}

Figure \ref{fig:beta7.5_planet} shows the spherically averaged gas density profiles as a function of distance from the planet centre for the simulations $\beta$7.5 and $\beta$7.5C9 for a few selected times. In both cases the planet contracts with time due to radiative cooling, however the contraction is understandably faster for the latter case. The planet is almost an order of magnitude denser at the end of the latter simulation, and its densest part is a factor of $\sim$ two more compact than that of the standard simulation $\beta$7.5. 

These numerical experiments show that the physical size of the planet and its contraction rate do matter in determining the accretion rate onto the planet. This is "unfortunate" but is a key part of the physical problem at hand. These physical uncertainties are larger than those due to numerical artefacts (\S \ref{sec:res}).
 
\subsubsection{Stochasticity due to interactions with the spiral arms}

\begin{figure}
\includegraphics[width=0.99\columnwidth]{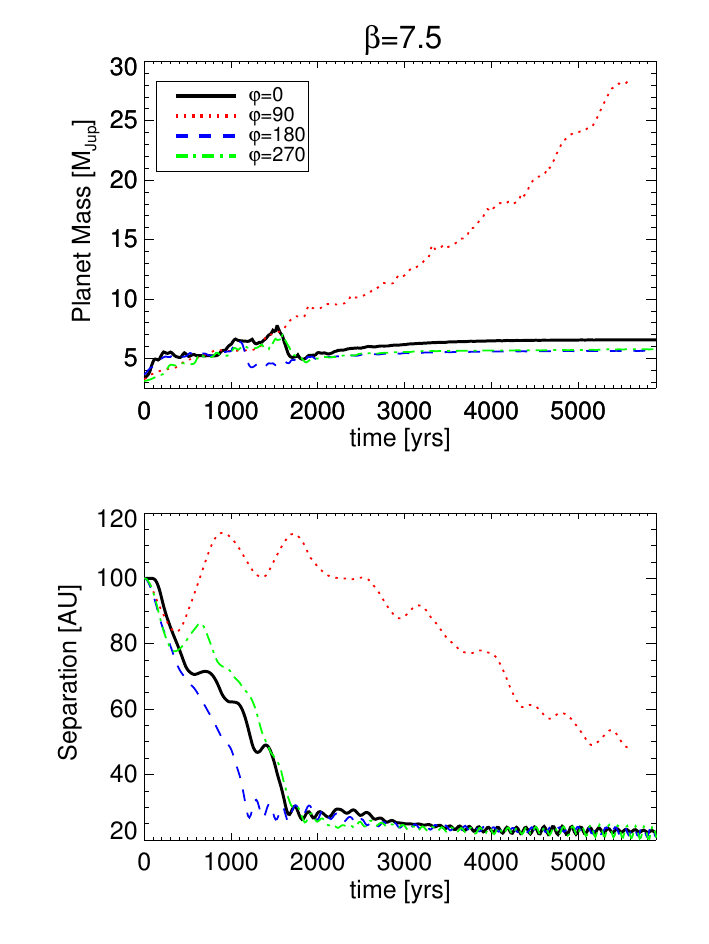}
 \caption{The planet accretion and migration tracks for the case $\beta=7.5$, differing only by the azimuthal angle $\phi$ for the planet starting position, as shown in the legend. The differences are driven by stochastic interactions with the spiral arms in the disc.}
   \label{fig:beta75}
 \end{figure}

Figure \ref{fig:beta75} shows how the results of the simulations in which the cooling rate parameter is fixed at $\beta=7.5$ change when the azimuthal angle of the planet initial position in the disc, $\phi$, is varied. The solid black curves are same as for the fiducial simulation $\beta7.5$, which was performed for $\phi=0$, whereas the three other sets of curves are for simulations in which $\phi = 90^\circ$, $180^\circ$ and $270^\circ$.

There is a significant dependence of the results on the angle $\phi$. This is consistent with the results of \cite{BaruteauEtal11} who found strong stochastic fluctuations in the planet migration patterns as they varied the starting $\phi$-angle of their planets. This is not surprising since the gas surface density and the velocity fields are highly non uniform  in strongly self-gravitating discs.

The starting angle $\phi=90^\circ$ yields the rests that differ from the rest the most. The planet reverses the direction of its migration in this case, wanders out to separation $R\sim 110-120$~au before migrating in. It takes the planet $\sim $ 3 times longer to eventually migrate into the inner disc than in the three other simulations displayed in the figure. The amount of material accreted by the planet in the $\phi = 90^\circ$ case is $\sim 25\mj$, which is an order of magnitude more than in the other cases. This clearly demonstrates that the length of the time available to the planet to accrete more gas and for that gas to contract to higher densities before it is plunged into the inner disc region, where the Hills radius is much smaller, is the key determinant of gas accretion onto the planet.

\subsection{Dependence of results on $\beta$}\label{sec:diff_beta}

\begin{figure*}
\includegraphics[width=0.9\columnwidth]{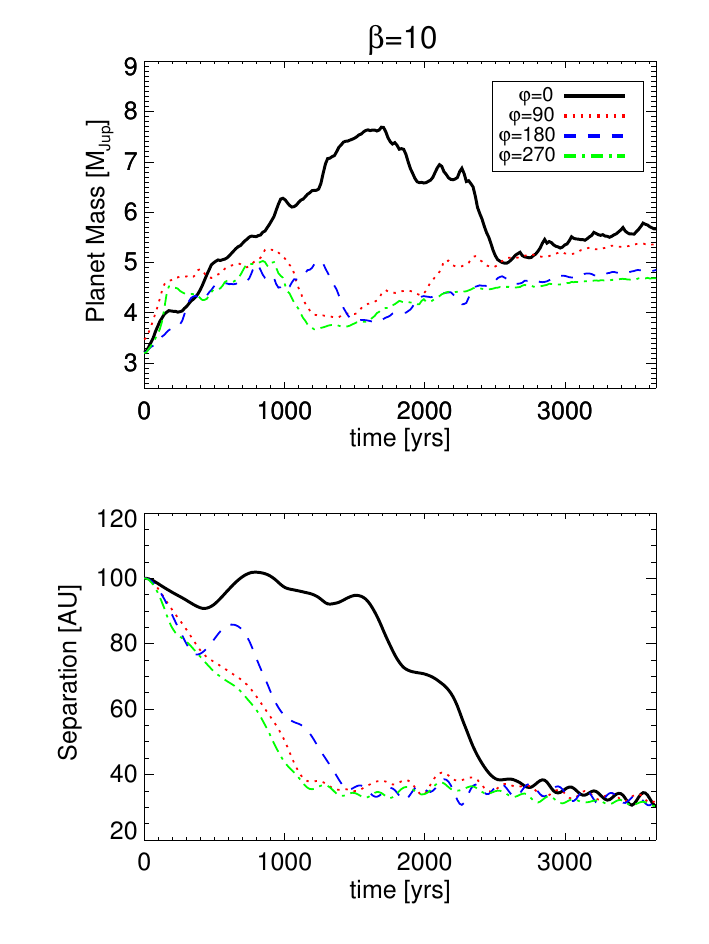}
\includegraphics[width=0.9\columnwidth]{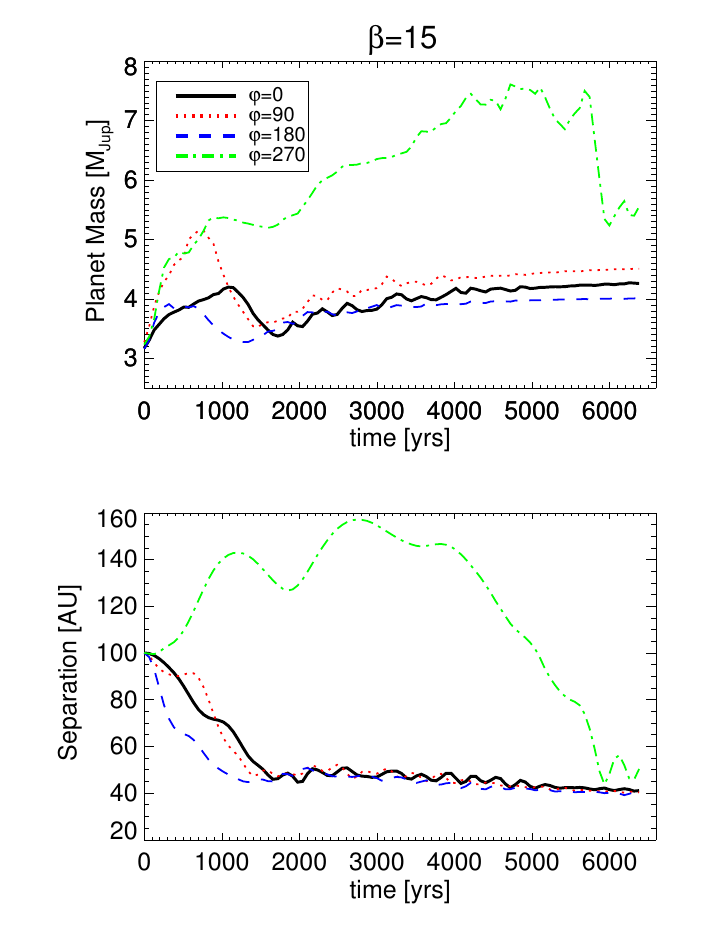}
\includegraphics[width=0.9\columnwidth]{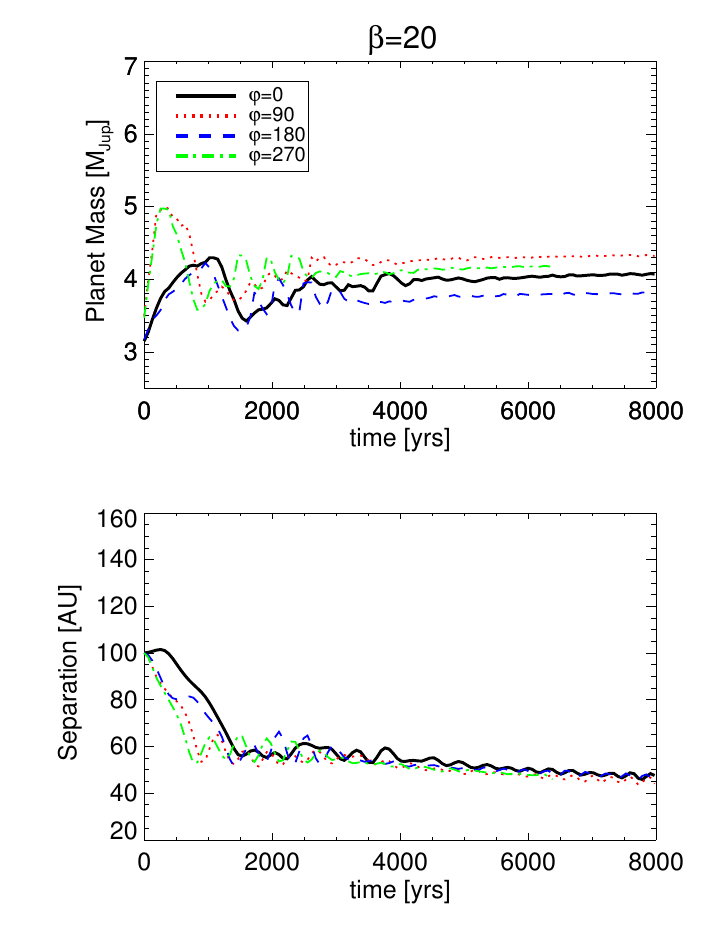}
\includegraphics[width=0.9\columnwidth]{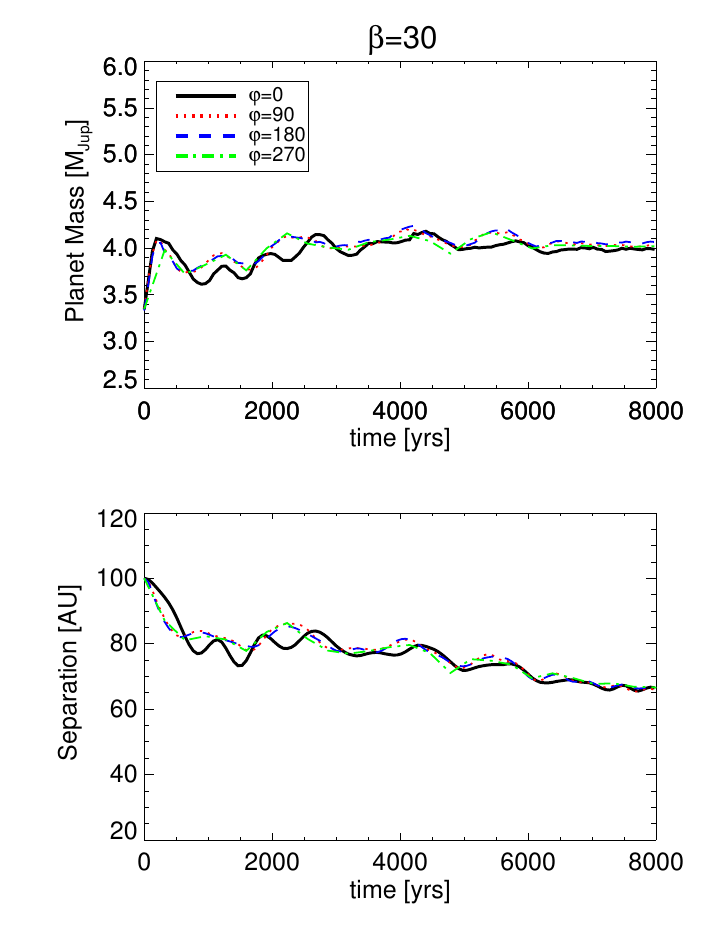}
 \caption{Same as Fig. \ref{fig:beta75}, but for four larger values of the cooling parameter $\beta$ as shown above the respective panels. Note a general trend: the larger the $\beta$, the less gas is accreted by the planet in general.}
   \label{fig:4beta}
 \end{figure*}

Fig. \ref{fig:4beta} shows the clump mass and separation versus time in similar calculations but for four larger values of the cooling rate parameter $\beta$. We see that the simulations for the two smaller values of $\beta$ (10 and 15) are qualitatively similar to the $\beta=7.5$ set of runs. In all of these three cases, one of the four starting values of $\phi$ led to the planet migrating outward first, or at least not migrating in as much (e.g., the case $\phi=0$ for $\beta=10$). In each of the simulations where the planet hang at wide separations for longer it managed to accrete much more gas than it did in the other runs with same value of $\beta$. This again indicates that the longer the period of time that the planet spends at wide separations before it migrates inward significantly, the more gas is accreted by the planet. Compared to the $\beta7.5\phi90$ case, the planet in the simulations $\beta10\phi0$ and $\beta15\phi270$ however loose much of the mass they accreted while at large separations when they migrate in rapidly. This can be understood by noting that it takes longer for the accreted atmospheres to contract at higher $\beta$, so the outer parts of the accreted atmospheres are "shaven off" when the planets plunge to smaller separations rapidly. In those cases planet accretion is a reversible process.

Another conclusion that can be drawn from figs. \ref{fig:beta75} and \ref{fig:4beta} is that the higher the value of $\beta$, the smaller the mass of the accreted gas onto the planet in general. In addition to that, the stochastic fluctuations in the planet migration tracks practically disappear for the two largest values of $\beta$. This is driven by the decrease in the density and mass of the spiral density arms in the discs with increasing $\beta$. In other words, the discs are much more uniform in the high $\beta$ cases.

\section{Discs with more realistic cooling}\label{sec:real}

Tests with the $\beta$-cooling model presented in the previous section reveal the importance of the gas cooling rate in determining the accretion rate onto the planet. However, this cooling model is unsatisfactory for a number of reasons. The cooling rate of real astrophysical flows must depend on the gas density, temperature, optical depth, and the presence of external irradiation (from the star, the planet and/or the larger scale parent molecular cloud). For the problem at hand this implies that the (effective) $\beta$ parameter changes as a function of separation from the star (usually decreasing with increasing separation), the distance from the planet, and the planet mass.

\subsection{The cooling model}

A simple yet widely used model for radiative cooling of discs is based on a plane-parallel geometry and an assumption that heat escapes via radiative diffusion \citep[e.g.,][]{Shakura73,Johnson03,Rafikov05,BoleyEtal10,VB10}. In particular, we follow \cite{GalvagniEtal12,NayakshinCha13}, and set the energy loss rate per unit volume as
\begin{equation}
\Lambda = \left( 36 \pi\right)^{1/3} {\sigma_{SB}\over  s} \left(T^4 - T_{\rm eq}^4 \right){\tau \over \tau^2 + 1}\;,
\label{coolr1}
\end{equation}
where $s$ is a length scale explained below and $\sigma_{SB}$ is the Stefan-Boltzmann constant. The equilibrium temperature $T_{\rm eq}$ is assumed to be established by the irradiation from the star and is set to
\begin{equation}
T_{\rm eq} = 20 \;\hbox{K } \left({R \over 100 \hbox{ au}}\right)^{-1/2}\;.
\label{Teq1}
\end{equation}
Most of the results presented below depend weakly on this choice for $T_{\rm eq}$.

The lengthscale $s$ is the disc scale height, and is approximated by $s = 0.1 R$. The disc optical depth is then estimated as 
\begin{equation}
\tau = \kappa \rho s\;,
\label{tau1}
\end{equation}
where $\rho$ is the local gas density and $\kappa$ is disc opacity. The latter is assumed to be dominated by dust, and is given by 
\begin{equation}
\kappa = f_{\rm op} \kappa_0(\rho, T)\;,
\label{kappa0}
\end{equation}
where $\kappa_0(\rho, T)$ is the interstellar dust opacity as given in Table 1 of \cite{ZhuEtal09}. The positive coefficient $f_{\rm op}$ is a free parameter, introduced to account for uncertainties such as grain growth, gas metallicity different from Solar, and other microphysical uncertainties in the dust opacity modelling, such as dust composition \citep[e.g.,][]{SemenovEtal03}.


\subsection{Initial conditions}\label{sec:IC}

The initial conditions for the simulations presented in this section are created by starting with a gas disc of mass $M_{\rm d} = 0.2 \msun$ with the surface density profile $\Sigma(R)\propto R^{-3/2}$ defined between the inner radius of 10 au and the outer radius of 140 au. The gas is set on the local circular orbits initially.  The initial number of SPH particles is $0.5$ Million, which corresponds to the SPH particle mass of $4\times 10^{-7} \msun$. There is no embedded planet in the disc, and the disc is relaxed for about 8,000 years using the cooling model explained above with $f_{\rm op} = 0.1$. With this relatively low value of $f_{\rm op}$, the disc cools down to about the equilibrium temperature (equation \ref{Teq1}) everywhere but is not self-gravitating due to its lower mass than in \S \ref{sec:beta}. The host star mass is initially $M_*=1\msun$, as previously, but its accretion (sink) radius is reduced to 2 au.


\subsection{Two contrasting cases}\label{sec:two_cases}

\begin{figure*}
\includegraphics[width=0.68\columnwidth]{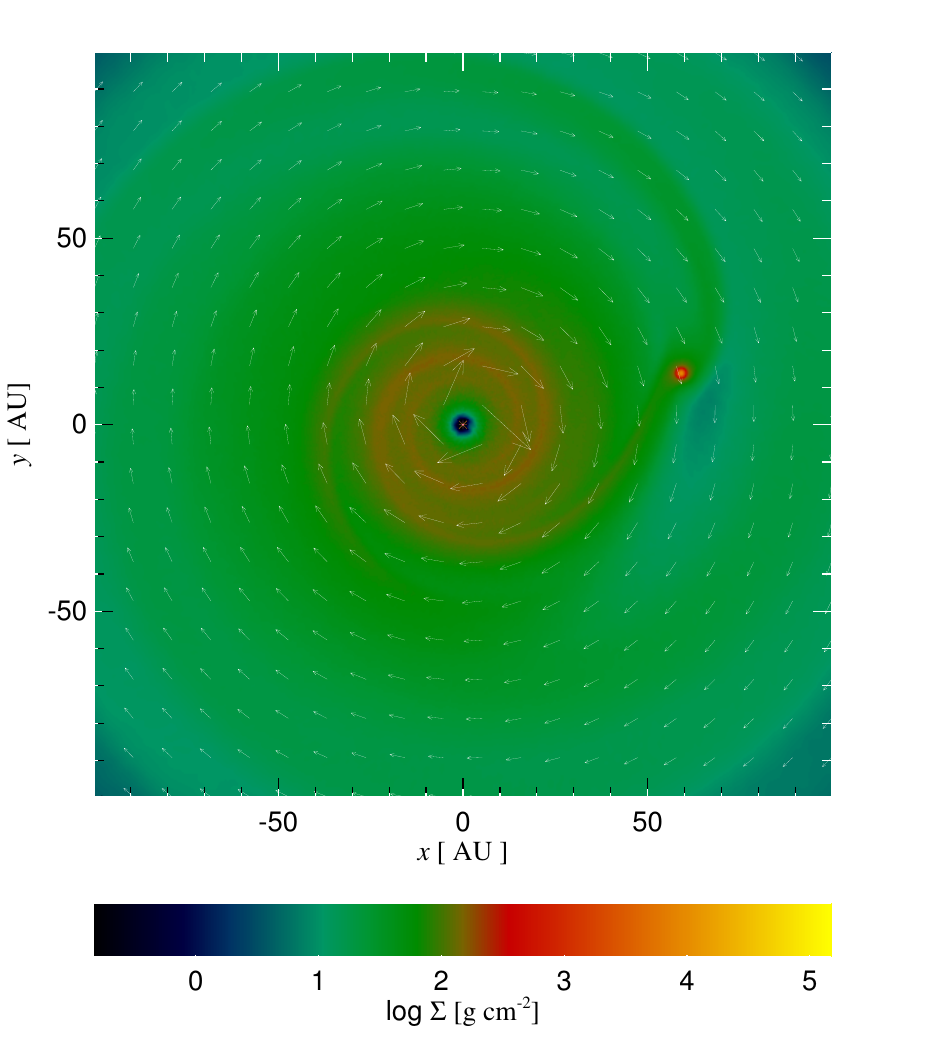}
\includegraphics[width=0.68\columnwidth]{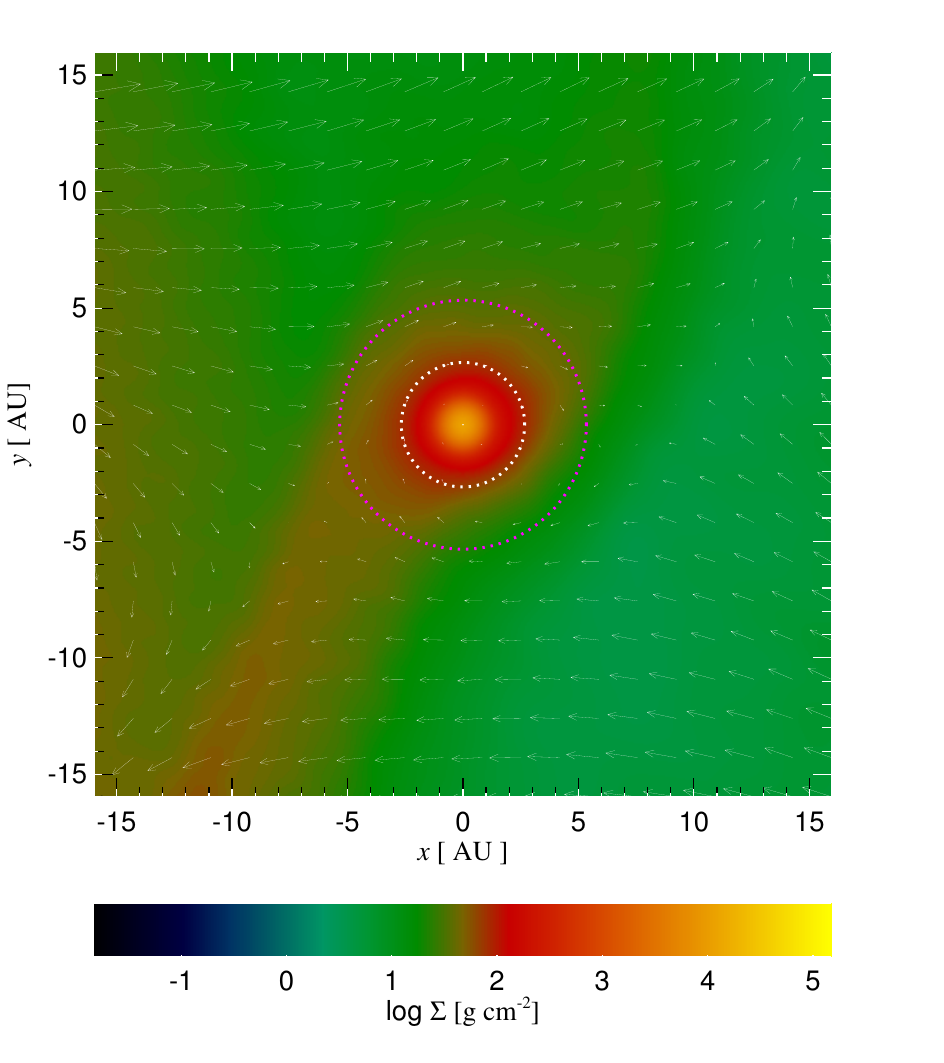}
\includegraphics[width=0.68\columnwidth]{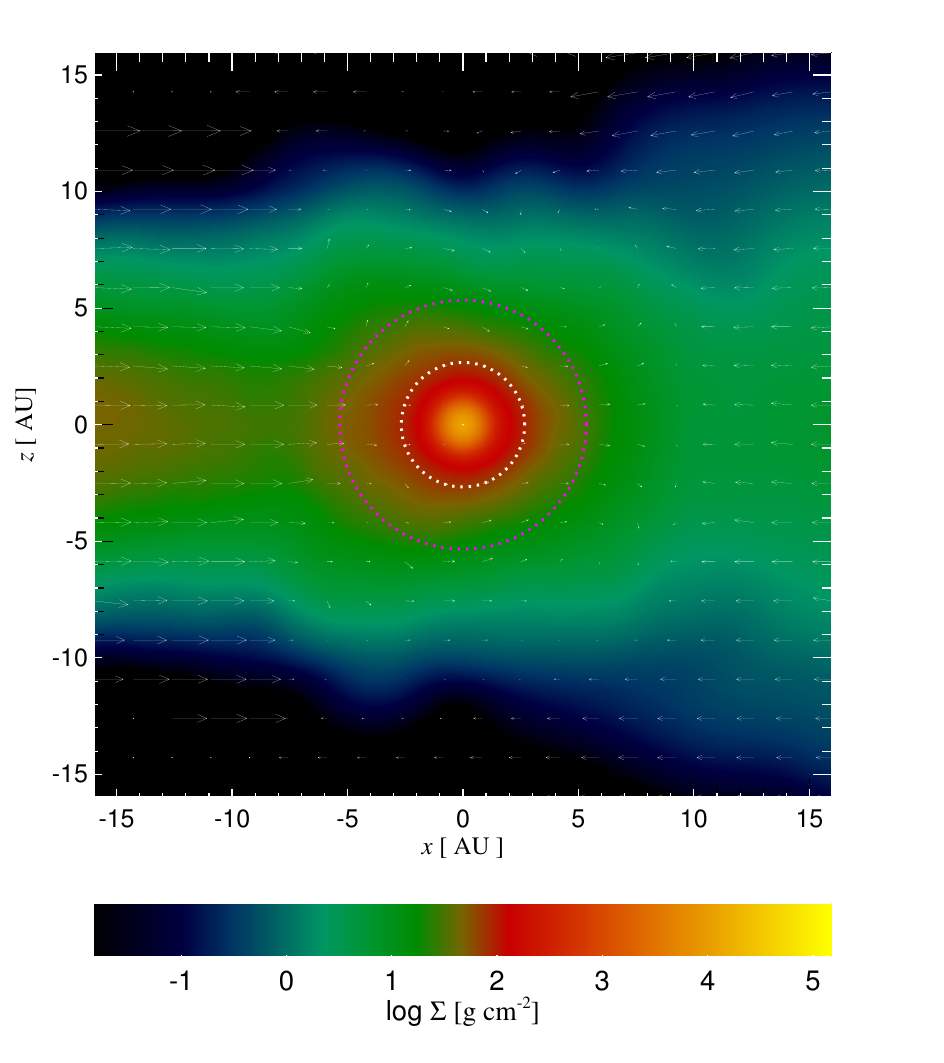}
 \caption{Snapshots from a simulation of a gas clump of initial mass of $2 \mj$ with opacity factor $f_{\rm op} =10$. The clump has migrated to separation $a=59$~au attracting hardly any gas from the surrounding disc. {\bf Left}: top view of the protoplanetary disc with the embedded clump. {\bf Middle}: zoom-in onto the clump location. {\bf Right}: same but a side-view projection of the region around the clump.}
   \label{fig:M2_f10}
 \end{figure*}

\begin{figure*}
\includegraphics[width=0.68\columnwidth]{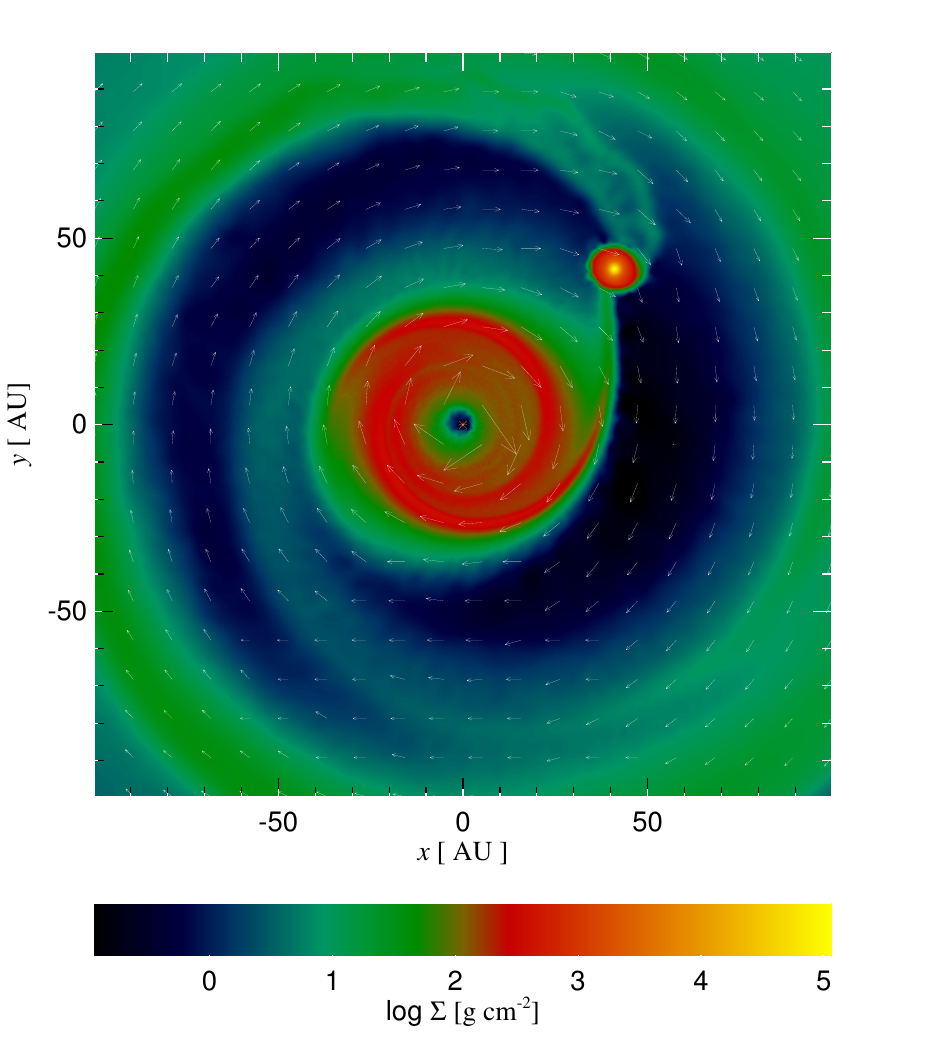}
\includegraphics[width=0.68\columnwidth]{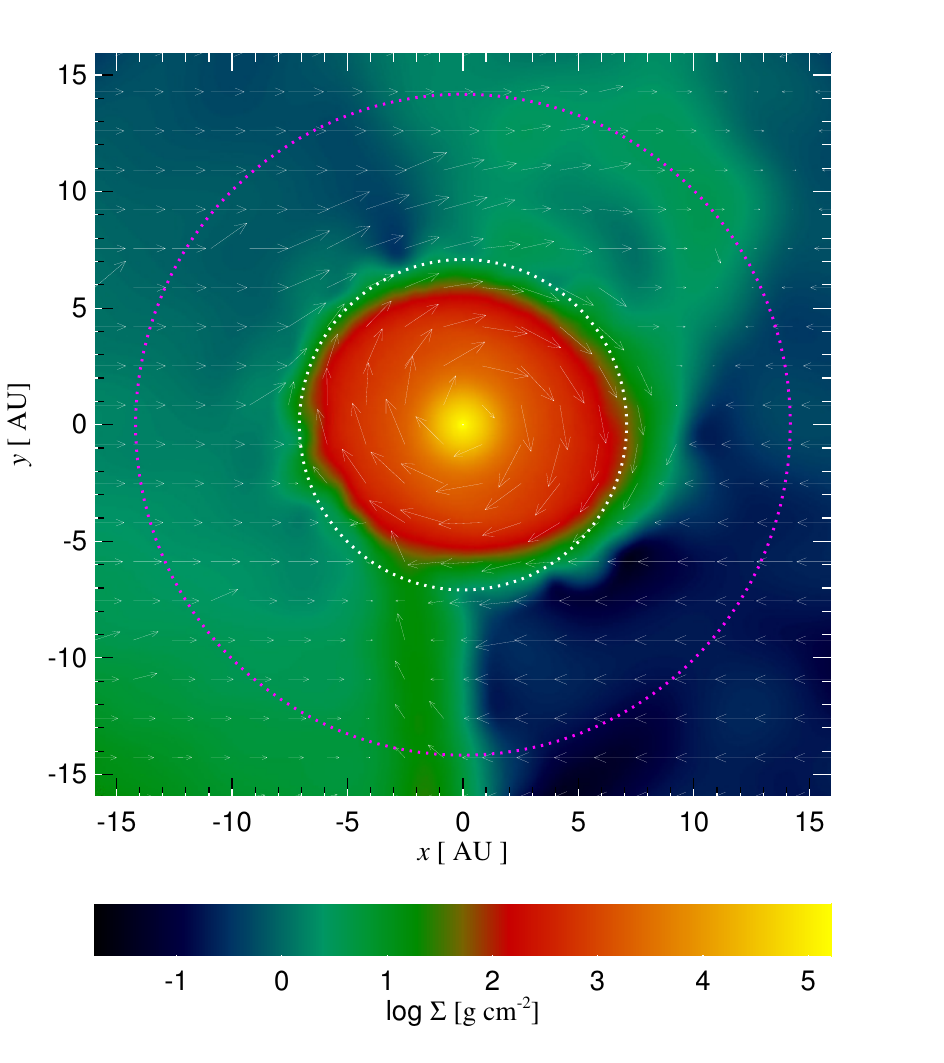}
\includegraphics[width=0.68\columnwidth]{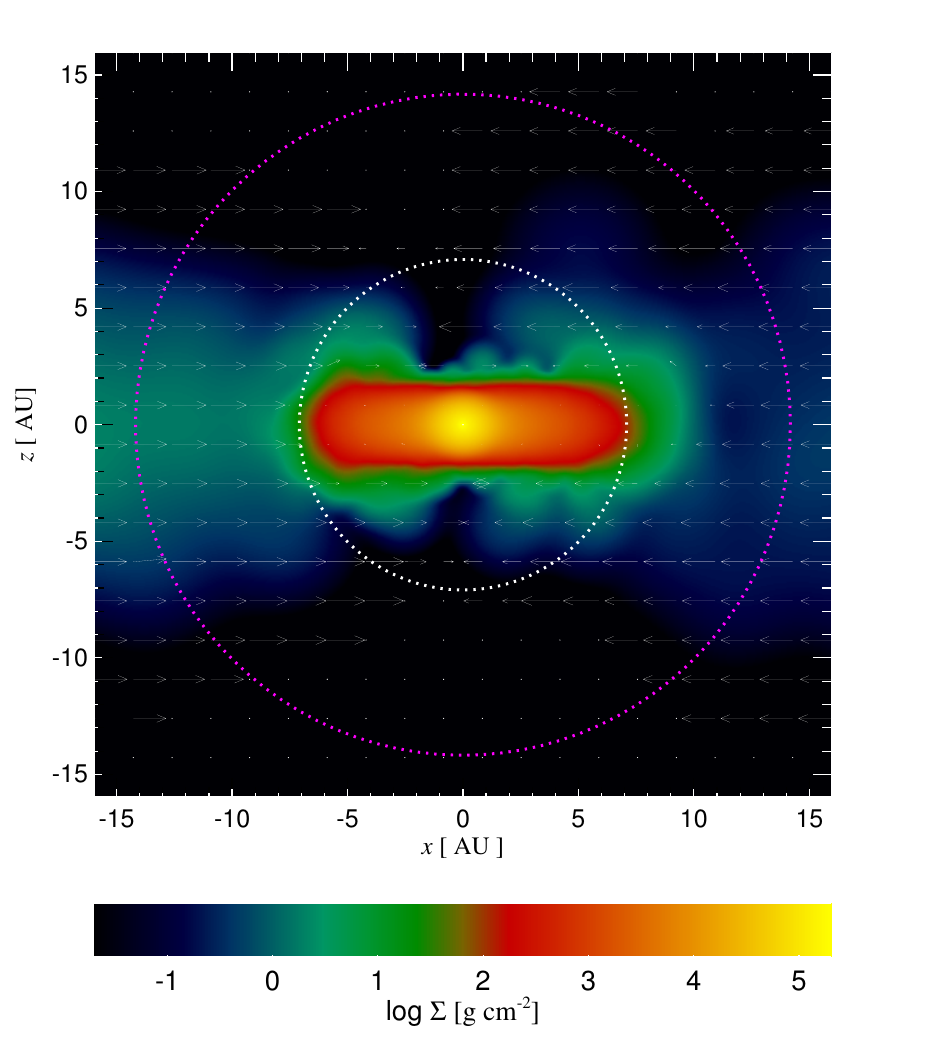}
 \caption{Same as fig. \ref{fig:M2_f10} but for the opacity factor of $f_{\rm op} = 0.01$. Due to the lower opacity, the clump accretes gas much more rapidly and hence becomes a massive brown dwarf, openning a wide gap in the parent disc. The contracting atmosphere around the clump has a rotationally supported disc geometry.}
   \label{fig:M2_f001}
 \end{figure*}

Figs. \ref{fig:M2_f10} and \ref{fig:M2_f001} show gas surface density maps from two simulations that are started from the same initial condition, with the initial planet mass of $2\mj$, but assume two very different opacity factors $f_{\rm op} = 10$ and $0.01$, respectively. 
For both of the figures, all of the panels are coeval, but the time is not the same between the two figures. The two cases were chosen to be compared at the time when the planets reached the same separation, $R = 59$~AU, rather than time. For the case of $f_{\rm op}=10$ this separation corresponds to time 2,630 years, whereas for the $f_{\rm op} = 0.01$ case the time is nearly exactly 4,000 years.

The left panels of the figures show the global disc structure in the face-on view. The middle and the right panels are centred on the planet location, and show the face-on and the edge-on views of gas flows around the planets, respectively.

There is a striking difference between the two cases. This is driven by the differences in the accretion rate onto the planets. In the $f_{\rm op} = 10$ case, the planet mass (defined as the total mass of gas within half the Hill sphere of the planet) is $M_{\rm p} = 2.06\mj$ at the time of the figure, hardly higher than the initial planet mass. The low opacity case, $f_{\rm op}= 0.01$, is a very different story -- here the planet mass is $M_{\rm p} = 43.06 \mj$. These differences reflect the fact that opacity strongly influences the rate at which the gas cools. Qualitatively, the high $f_{\rm op}$ case reflects the very long cooling time (large $\beta$) regime, whereas the low $f_{\rm op}$ case corresponds to the short cooling time cases (small $\beta$). 

The high opacity case planet, remaining too low mass, continues to migrate in the type I regime \citep{BaruteauEtal11}, affecting the disc structure around it much less than the low opacity case planet which actually rans away well into the brown dwarf mass regime. A deep broad gap is opened in the gas disc in this case (fig. \ref{fig:M2_f001}). The planet mass is about half of the isolation mass \citep{Lissauer87}, $M_{\rm iso} \sim M_{\rm d} \sqrt{M_{\rm d}/M_*} \approx 0.09 \msun$, suggesting that the growth of the planet cannot go on at this high rate for much longer.  Due to the gap and also due to its large mass, the planet migrates inward slowly in this regime.

The morphology of gas around the planets is also different in these two cases. In the high opacity case, the atmosphere around the planet is quasi-spherical, being hot and weakly bound to the planet. Note how small velocity vectors become inside the half $R_{\rm H}$ radius of the planet, suggesting that all of the support against gravity there is due to thermal gas pressure. In the opposite case of $f_{\rm op} = 0.01$, the thermal pressure support is minimal because the radiative cooling of the gas captured by the planet is rapid and so the gas is sub-virial inside the Hill sphere. The Hill sphere of the planet is also larger due to the high mass that the planet achieved by the time of the snapshot. For this reason the angular momentum of the gas captured by the planet from the disc is larger in fig. \ref{fig:M2_f001}, so the gas settles into a disc which is supported mainly by rotation around the planet rather than thermal pressure.

 These two contrasting cases show that the rate of cooling remains the key determinant of what happens to a clump embedded in a massive protoplanetary disc in the more realistic although approximate radiative cooling model for the disc. As with $\beta$-cooling discs, the longer the cooling rate, the less efficient the gas accretion on the planet is.

\begin{figure*}
\includegraphics[width=0.99\columnwidth]{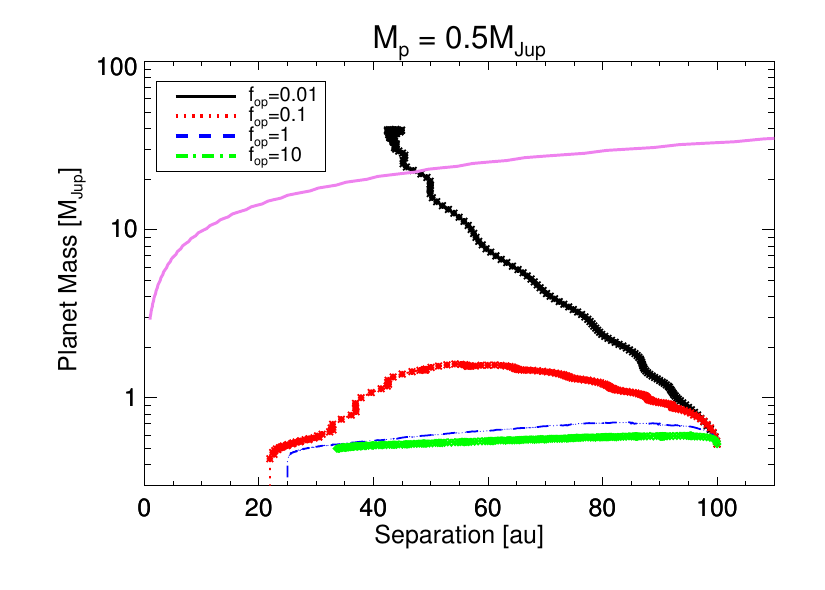}
\includegraphics[width=0.99\columnwidth]{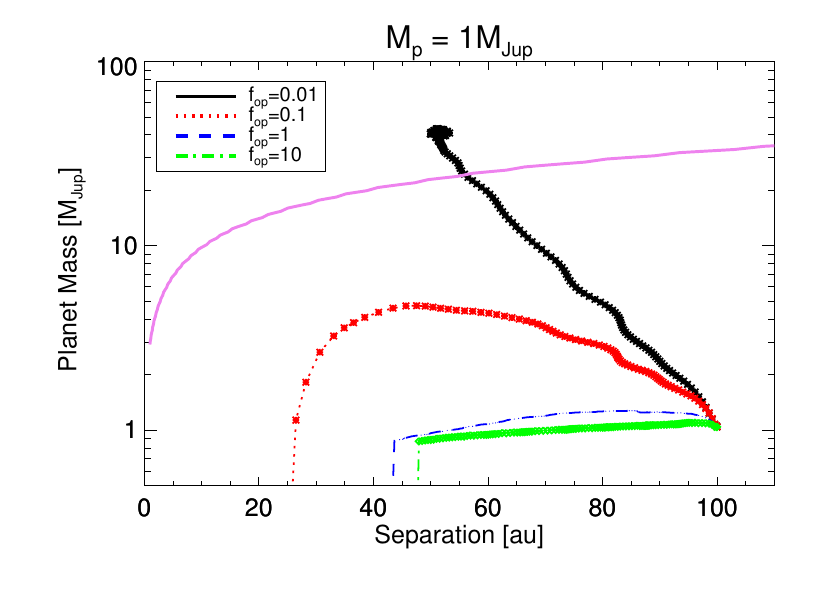}
\includegraphics[width=0.99\columnwidth]{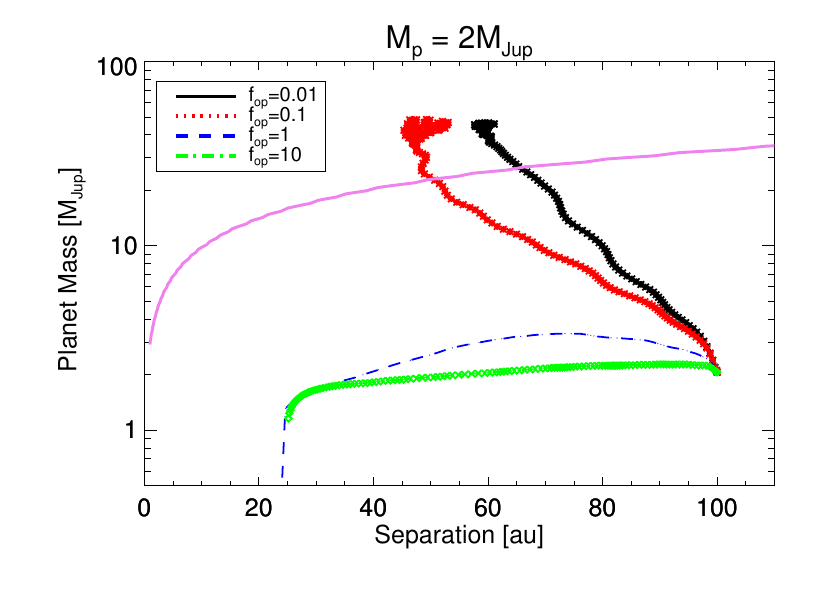}
\includegraphics[width=0.99\columnwidth]{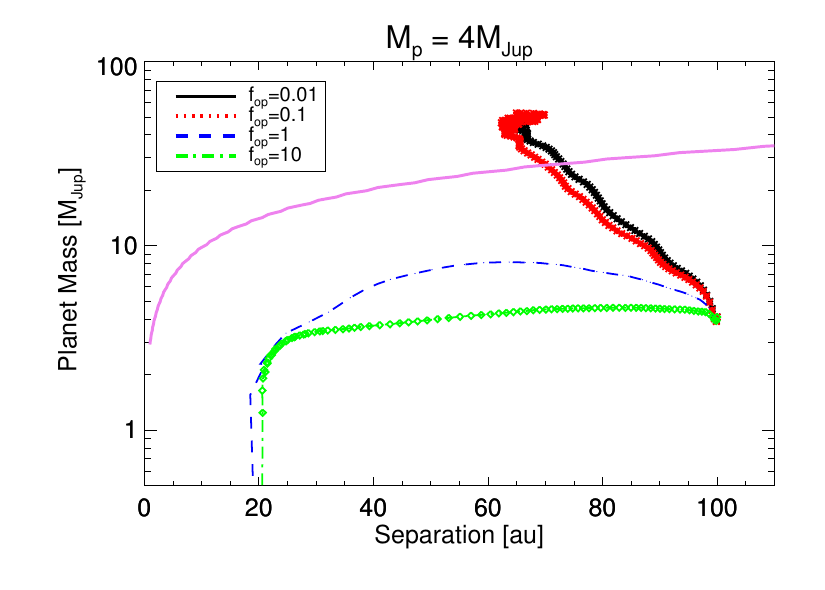}
\includegraphics[width=0.99\columnwidth]{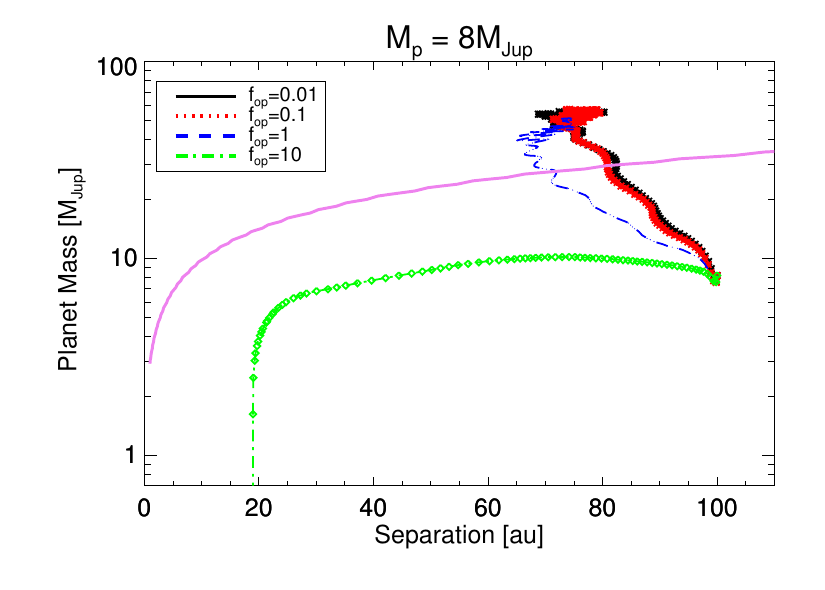}
\includegraphics[width=0.99\columnwidth]{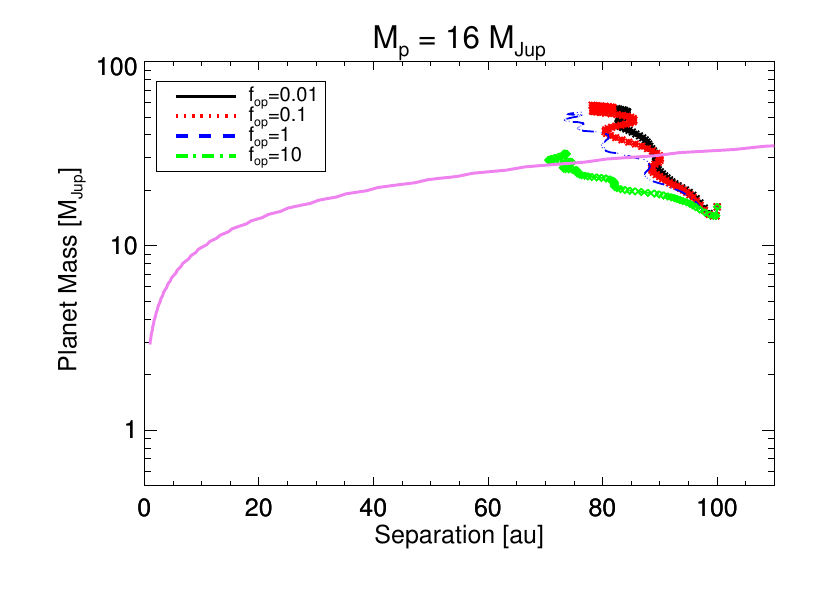}
 \caption{Mass versus separation for pre-collapse gas clumps of different masses, all starting on a circular orbit at $a=100$~au initially. The violet colour curve is the planet mass needed to open a deep gap in the disc (eq. \ref{Crida}), and it is the same in all the panels. In each panel, the lines of different style refer to the different opacity factor $f_{\rm op}$, as explained in the legend, which controls the cooling rate. For low $f_{\rm op}$, gas cools rapidly, fuelling rapid gas accretion onto the planet. For large $f_{\rm op}$, gas accretion is inefficient. See \S \ref{sec:suit1} for detail.}
   \label{fig:Tracks}
 \end{figure*}

\subsection{A suit of simulations}\label{sec:suit1}

Fig. \ref{fig:Tracks} shows the mass-separation tracks for a number of simulations with set-up and initial conditions identical to those explained above, but now for a range of the initial planet mass and several values of $f_{\rm op}$. In particular, the initial masses explored here are $0.5$, 1, 2, 4, 8 and $16 \mj$; the opacity factor $f_{\rm op}$ is set to $0.01$, $0.1$, 1 and 10. The vertical axis shows the clump mass in Jupiter masses while the horizontal axis shows the clump separation in au. The violet (parabola-like) curve is the gap-opening planet mass, $M_{\rm gap}$, plotted versus radius. It is defined following \cite{CridaEtal06}, who have found that parameter
\begin{equation}
C_{\rm p} = {3\over 4} {H\over R_{\rm H}} + {50 \alpha H^2\over R^2} {M_*\over M_{\rm p}}\;
\label{Crida}
\end{equation}
needs to be smaller than unity for the planet to open a gap. Here $\alpha$ is the viscosity parameter which we set to $\alpha = 0.06$ in this equation, which is appropriate for strongly self-gravitating discs \citep{Rice05}. In computing $C_{\rm p}$, the disc vertical scale height is found from $H/R = (k_B T_{\rm eq} R/G M_* \mu)^{1/2}$, where $T_{\rm eq}$ is the equilibrium irradiation temperature given by eq. \ref{Teq1}.

The figure shows that the planets grow in mass rapidly, becoming massive brown dwarfs, if dust opacity is low, e.g., for $f_{\rm op} \ll 1$, when the radiative cooling is rapid. This result is particularly striking for the planet with the initial mass of $0.5 \mj$ (black solid track in the top left panel), which increased in mass by a factor of about 100 during the simulation with $f_{\rm op} = 0.01$, but migrated in only by a factor of $\sim 2$. As explained previously, the physical reason for the very massive objects to not migrate much in their discs during these simulations is the gap that they open in the disc. Due to that gap, they switch to the slower type II migration. 

Slowly cooling planets, that is high dust opacity cases (green and blue curves), however, do not grow in mass significantly except for the most massive planet case. In this limit the planets are unable to accrete gas rapidly and so continue to migrate in quickly in the type I regime. As a result, they eventually migrate "too far", where their Hill radius is about the physical size of the planet, so that the planets become unbound. 

Comparing now the planet migration tracks to the $M_{\rm gap}(R)$ curve, we note that the curve does a reasonably accurate job at predicting when the migration rate of a planet drops significantly (this will also be discussed further in \S \ref{sec:post}). Indeed, once a planet runs away towards masses of $\sim 20-30 \mj$, the migration tracks become approximately vertical, implying a much slower migration rate. Additionally, planet tracks get "smudged" in Fig. \ref{fig:Tracks} when the planets open deep gaps because the planet spends in that nearly stalled location more time. Some eccentricity growth is also apparent after the planets open gaps. In comparing the migration tracks to equation \ref{Crida}, it should also be noted that this equation is derived assuming a planet on a fixed orbit. \cite{MalikEtal15} have recently showed that a gap opening planet must have a migration time scale which is shorter than the timescale for the disc viscous flow to close the gap, which makes gap opening more difficult, increasing the estimate above $M_{\rm gap}$ somewhat. 

A secondary effect can be glimpsed from the figure: the higher the initial mass of the clump, the less likely it is to be tidally disrupted and the more likely it is to end up as a massive brown dwarf. The highest initial mass case, $M_{\rm p} = 16\mj$, is not disrupted and grows more massive for all of the values of $f_{\rm op}$ considered. In contrast, for the lowest initial planet mass, $M_{\rm p} =0.5\mj$ case, only the smallest $f_{\rm op} = 0.01$ clump "runs away" into the brown dwarf domain, with the three higher opacity cases leading to tidal disruption.

This sensitivity to the initial planet mass is due to more massive gas clumps contracting much more rapidly\citep{Nayakshin10a,HB11,Nayakshin15a}. As we have seen in \S \ref{sec:size}, the smaller the physical size of the planet, the larger is the accretion rate on it. 

 
 \section{Accretion on post-collapse planets}\label{sec:post}
 
 \subsection{Using sinks for gas accretion}\label{sec:sinks}
 
 So far only the pre-collapse gas clumps were considered in the study. That is a very important case to consider. However, there may be processes not included or not resolved in our study that allow gas fragments to contract and then collapse more rapidly than our simple radiative cooling models allow. For example, grain growth may be much more rapid inside the planet than in the disc since the planet is much denser than the disc \citep[e.g.,][]{ChaNayakshin11a}. If grain opacity of the planet is reduced by grain growth sufficiently, then the planet may collapse much faster \citep[e.g.,][]{HB11} than our cooling prescriptions allow. Post-collapse planets are multiple orders of magnitude denser than the pre-collapse ones. Processes such as small scale convection inside the planet must be resolved in detail in order to follow their evolution on Million year or longer time scales carefully \citep[e.g.,][]{BurrowsEtal01}, which is clearly not possible  here due to numerical simulations.
 
There is therefore little choice but to resort to the sink particle approach \citep{Bate95} to model post-collapse planets in 3D simulations of protoplanetary discs.  The simplest capture-all within sink radius $r_{\rm sink}$ approaches creates an artificial vacuum in that region and an unphysical pressure gradient force, almost certainly leading to over-estimating accretion rate unless $r_{\rm sink}$ is very (usually numerically uncomfortably) small \citep[e.g.,][]{Cuadra06}. To prevent the development of the unphysical positive pressure gradient at $r\approx r_{\rm sink}$, I soften the gravitational potential of the planet with $h_{\rm pl} = 0.2$~AU. The sink radius for the planet used for simulations in this section is set to $r_{\rm sink} = 0.3$~AU, which is much smaller than the sink radius for the star (set to to 5 au here). 

In addition, only the gas particles that are (a) inside distance $r_{\rm sink}$ from the planet/sink and (b) whose gas density exceeds $\rho_{\rm cr} = 5\times 10^{-12}$ are accreted. Numerical experiments showed that these conditions prevent accretion of hot particles not physically bound to the sink. 
Gas captured by the planet's potential forms a dense "atmosphere" around the planet which results in development of a negative pressure gradient, as physically expected. This atmosphere prevents particles from the disc entering the Hills sphere of the planet only temporarily from being artificially accreted. The newly arrived gas gets bound to the planet only if it cools rapidly. In that case the mass and the density of the atmosphere around the sink increases with time and eventually both of the above conditions for gas accretion are satisfied. This approach therefore provides a barrier and a time delay in accretion of gas particles onto the sink and hence differentiates between the physical and unphysical accretion of gas onto the sinks.

\subsection{Results}\label{sec:sink_results}

The initial conditions for the disc and the planet used in this section are identical to those used in \S \ref{sec:real} except that sink particles are used to model the planets. The planet mass measured in the runs presented here is the sink mass rather than the mass within half of the Hills sphere (as in \S \ref{sec:real}). This does not lead to noticeable differences in results since for post-collapse planets the sink accounts for nearly all mass within the Hill's sphere anyway.

The suit of numerical experiments with the initial planet mass ranging from $0.5 \mj$ to $16\mj$ and the disc opacity factor varied from $f_{\rm op} = 0.01$ to 10 (all as in \S \ref{sec:real}) is then run.

\begin{figure}
\includegraphics[width=0.99\columnwidth]{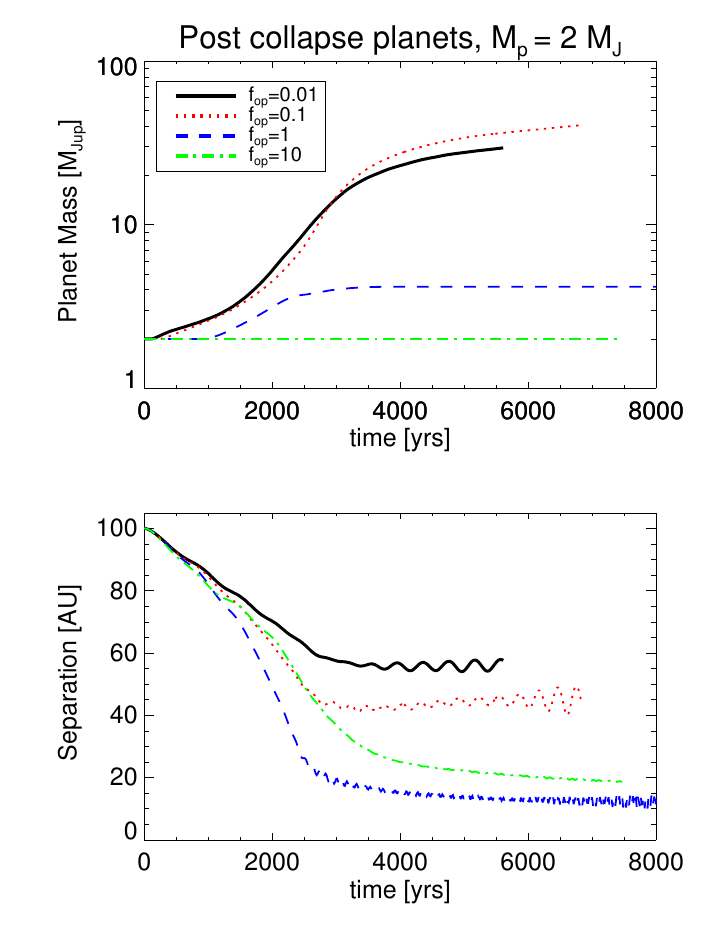}
 \caption{Planet mass (top panel) and separation (lower panel) versus time for post-collapse planet simulations with starting mass $M_{\rm p} = 2 \mj$. As before, low opacity cases accrete gas rapidly and run away to high masses whereas high opacity cases do not accrete gas and instead migrate in at a fixed mass (see \S \ref{sec:post} for discussion).}
   \label{fig:Accr_sink}
 \end{figure}

 Figure \ref{fig:Accr_sink} shows simulations performed for the initial planet mass of $M_{\rm p} = 2 \mj$. We observe that, as previously, low opacity cases result in a rapid gas accretion onto the planet, so that it runs away to brown dwarf masses. Larger values of $f_{\rm op}$ (dashed and dot-dashed curves) result in the planet accreting less or no gas at all. These planets migrate in faster and further because they do not open a gap in the parent disc. Also note that when the planets open deep gaps in the discs they slow down and so this is where the separation versus time plot becomes nearly horizontal in Fig. \ref{fig:Accr_sink}. For the two low opacity cases, $f_{\rm op} = 0.01$ and 0.1, this occurs at around mass $M_{\rm p} \approx 20\mj$, when the planets are at separation $\sim 40-60$~au. This is very close to what equation \ref{Crida} predicts  for $M_{\rm gap}$ at such separations (cf. Fig. \ref{fig:Tracks}). For the two higher opacity cases, the planets stall at separation $\sim 10-20$~au. Their masses are only 2 and $4\mj$ at that point, which is a factor of $\sim 2-3$ smaller than eq. \ref{Crida} predicts. However, the sink radius of the star is 5 au in these simulations, implying that the structure of the disc at around 10 au is probably significantly affected by it. It is likely that this makes gap opening somewhat easier,
 explaining why the planets tend to stall near the inner edge of the disc even if their mass is somewhat below $M_{\rm gap}$. Additionally, the lower mass of the disc near the inner edge also means that planets migrate in slower than they would if the flow was better resolved to much smaller radii.

 \begin{figure*}
\includegraphics[width=0.99\columnwidth]{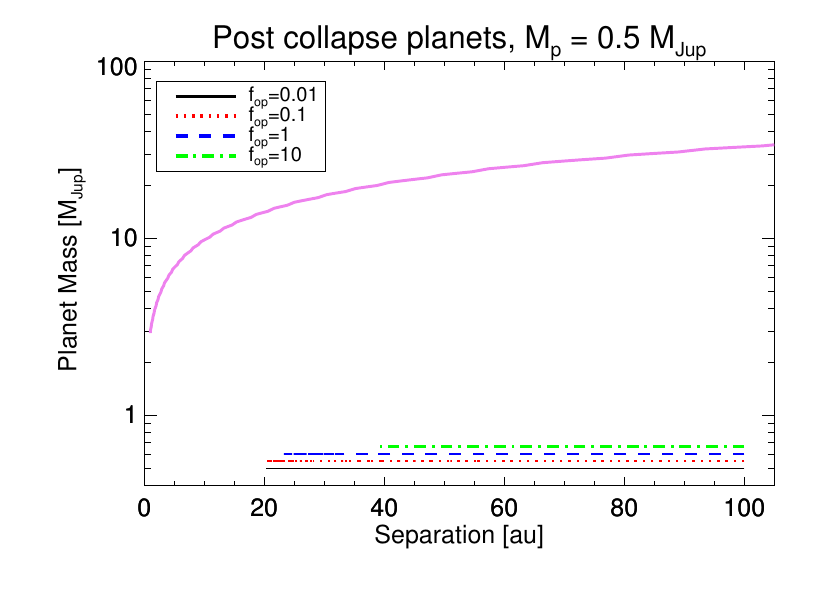}
\includegraphics[width=0.99\columnwidth]{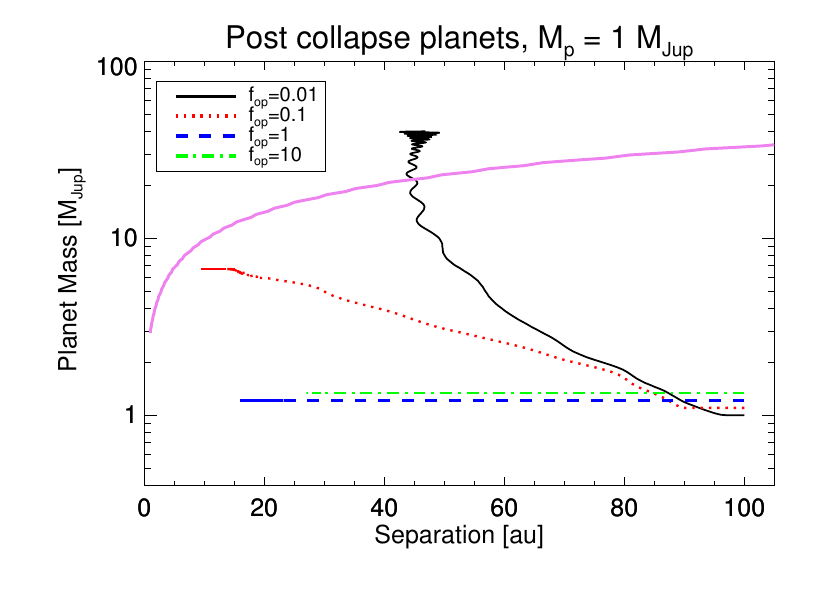}
\includegraphics[width=0.99\columnwidth]{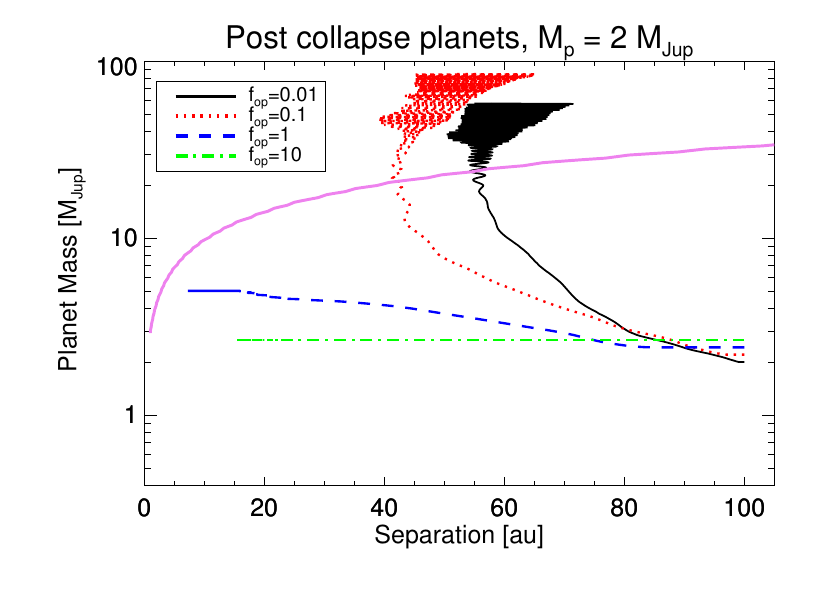}
\includegraphics[width=0.99\columnwidth]{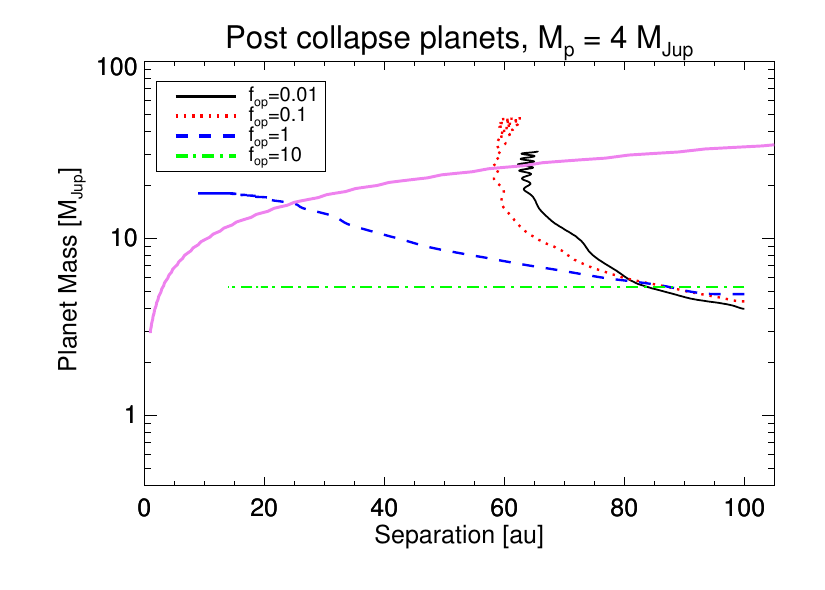}
\includegraphics[width=0.99\columnwidth]{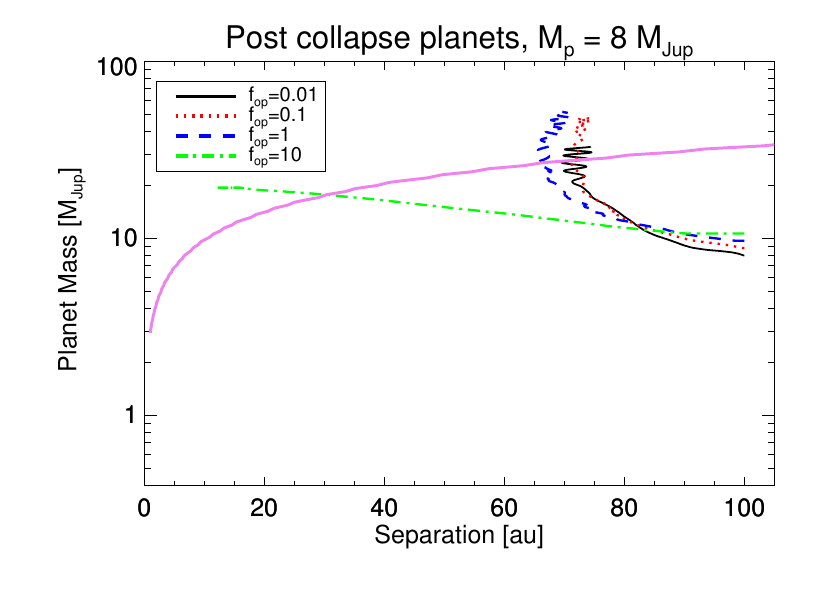}
\includegraphics[width=0.99\columnwidth]{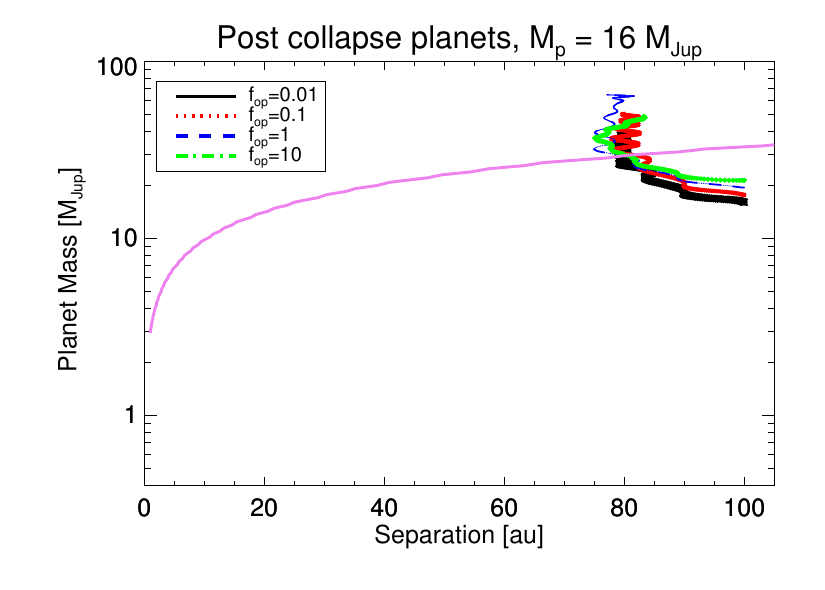}
 \caption{Same as Fig. \ref{fig:Tracks} but for post collapse planets modelled as sink particles. As previously, planets accrete gas rapidly when the opacity factor $f_{\rm op}$ is low, allowing gas to cool rapidly. In the opposite case planet growth stalls and then migrate inward quickly instead. Since post collapse planets are very dense, they are not disrupted tidally in the simulations even if they migrate to the inner boundary of the computational domain.}
   \label{fig:Tracks_sinks}
 \end{figure*}
 
 Fig. \ref{fig:Tracks_sinks} shows the same grid of post-collapse planet simulations as was shown in Fig. \ref{fig:Tracks} for pre-collapse planets. To improve visibility of the similar curves in Fig. \ref{fig:Tracks_sinks}, the curves are shifted by the factor of 1.1 vertically from one another.
 
The results of the sink particle runs are overall very similar to those obtained in \S \ref{sec:real} as far as gas accretion is concerned. We again notice that low opacity cases (low $f_{\rm op}$) encourage much faster accretion of gas onto the planets. We also note that there is a trend with the planet initial mass. The higher the mass of the planet, the more likely it is to run away into the massive brown dwarf territory by accreting gas rapidly. Minor differences do appear, such as the fact that none of the post-collapse runs for the lowest planet mass, $M_{\rm p} = 0.5\mj$, resulted in a massive object, whereas the $f_{\rm op} = 0.01$ run from the respective panel in Fig. \ref{fig:Tracks} did runaway towards a brown dwarf mass. This minor difference may be sensitive to the sink particle prescription parameters, such as the sink radius.

The largest difference between the post-collapse case and the pre-collapse calculations is that in the former case the planets are a factor of $\sim 100$ more compact and are therefore not tidally disrupted. The end result of the runs that do not accrete gas in fig. \ref{fig:Tracks_sinks} is a gas giant planet that migrated to the inner $\sim 10-20$~au. We would need to follow these cases right to the moment of their disc dissipation to ascertain the outcome, but the range of outcomes is quite clear. If the disc is removed rapidly then these giant planets may be cold or warm gas giants (with separation from sub au to $\sim 10$~au). If the disc is removed later then the planet may become a hot Jupiter or even be plunged all the way into the star.

 \section{Discussion}\label{sec:discussion}
 
\subsection{Main Results}\label{sec:main} 

\begin{figure*}
\includegraphics[width=0.99\columnwidth]{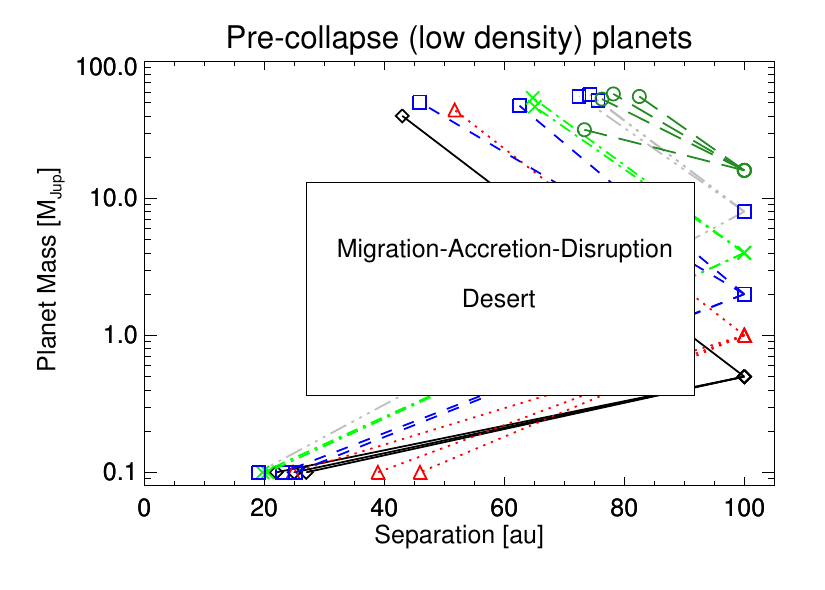}
\includegraphics[width=0.99\columnwidth]{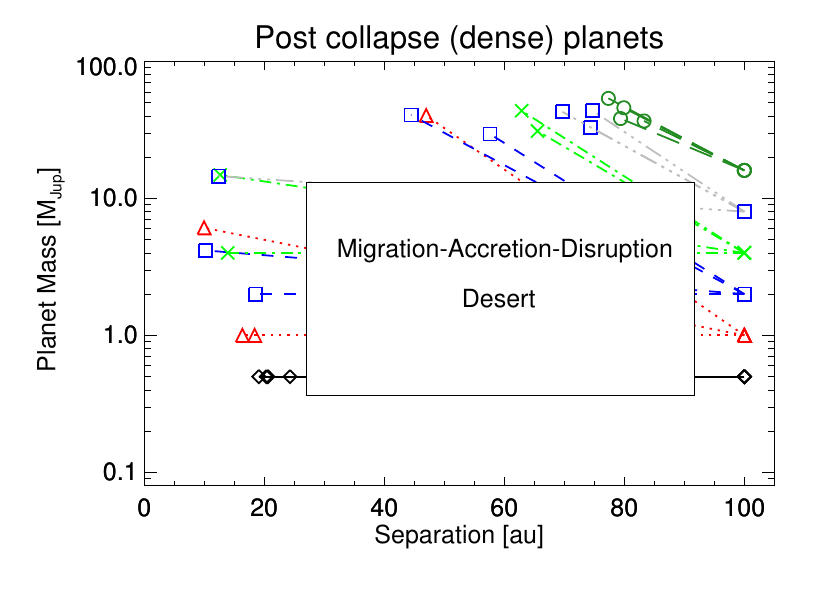}
 \caption{The initial and the final positions of the simulated planets in the mass versus separation parameter space. For each simulation, the initial mass and planet position is connected with the final quantities by a straight line. Note that not a single simulation ended up within the boxed region which is termed a desert. The desert is due to the clumps being taken out of that region by the inward migration, gas accretion or tidal disruption of pre-collapse planets. In the latter case it is assumed that the clumps made massive solid cores which remain behind after the disruption of the clump. The mass of the remaining core (plus any envelope if it retains one) is set to $0.1\mj$ arbitrarily for clarity of the figure.}
   \label{fig:MoneyPlot}
 \end{figure*}

Fig. \ref{fig:MoneyPlot} summarises the results of this paper. The panels on the left and on the right show the pre-collapse planets (studied in \S \ref{sec:real}) and the post collapse ones (\S \ref{sec:post}), respectively. Each line in the plot represents one simulation and connects the initial planet mass and separation with the final ones. The different initial planet mass cases are plotted with different symbols and line styles and colours. The lines of the same types differ by the opacity parameter $f_{\rm op}$. For example, for the four dotted red lines, starting from the triangle at initial values $(a,M_{\rm p}) = $~(100 au, $1 \mj$) in the left panel, three simulations with $f_{\rm op} = $~10, 1 and 0.1 led to the fragment tidal disruption at a few tens of au separation (cf. the top right panel in fig. \ref{fig:Tracks}), whereas the $f_{\rm op}= 0.01$ simulation resulted in the planet becoming a massive brown dwarf by accreting gas rapidly. 

We can see that where the clump will end up in the mass-separation phase space in the end of a simulation depends very sensitively on the opacity of the run, which unfortunately remains very uncertain for protoplanetary discs. However, there is one opacity independent outcome for all of the simulations performed in this paper. There is a desert of gas giant planets with separation greater than $\sim 20$~au (more realistically $\sim 10$~au, see below). The desert is depicted by a rectangular box in the figure. The desert existence is due the trio of processes -- migration, accretion and disruption -- always removing the gas clumps from that region. Variations in the input physics or parameters of the simulation control which of the three processes is dominant in removing the gas giants from the desert. This therefore affects where a gas clump starting at $a=100$~au ends up in the diagram, but all of the possible outcomes from the simulations presented here are outside the desert. 

If gas clumps are in the pre-collapse (youngest) state, hydrogen is molecular, and the clump density is $\sim 7$ to $10$ orders of magnitude lower than that of the present day Jupiter \citep[e.g.,][]{Nayakshin10a}. The clump is then found to evolve in one of two ways.

{\em Disruption}. If gas opacity is high, then the cooling rate of gas captured into the Hills sphere of the planet is low, and hence gas accretion onto the clump is inefficient (see analytical arguments in \S \ref{sec:analytics}). The clump also does not contract significantly for the same reason. It migrates inward at more or less constant mass until it fills its Roche lobe and gets disrupted by the tidal forces from the host star \citep{VB06,BoleyEtal10}. Gas sedimentation inside the clump \citep{McCreaWilliams65,Boss97,HS08,HelledEtal08,Nayakshin10a,Nayakshin10b}, not taken into account here, may form massive solid cores which are left behind after the gas clump is disrupted \citep{BoleyEtal10,Nayakshin10c,ChaNayakshin11a}. These massive cores are shown as symbols below the box in the left panel in fig. \ref{fig:MoneyPlot}. The mass of the core is arbitrarily set at $0.1 \mj$ in the figure but can be significantly less than that, especially in metal-poor environments.

{\em Runaway accretion into the brown dwarf phase}. When the gas opacity factor is low, gas accretion onto the clump is generally rapid. The clump grows in mass in a runaway fashion until it opens a gap in the disc. The clump exits the desert box through the upper boundary, becoming a massive brown dwarf. Its further evolution depends on the disc evolution on $\sim 1$ Million year time scales, not modelled here, and is hence not clear. It could become even more massive by accreting more gas from the disc and/or it may migrate inward, potentially all the way in the inner 1 au from the host star. 

On the other hand, grain growth inside the gas clumps or metal loading onto the clumps, both not studied here, may allow the clumps to collapse much faster \citep{HB11,Nayakshin15a}. In this case we should focus on the simulations with post collapse planets, studied in \S \ref{sec:post}, the results of which are shown in the right panel of fig. \ref{fig:MoneyPlot}. These planets are much denser, with typical central densities $\sim 10^{-3}$ g~cm$^{-3}$ or higher. These planets are too dense to be tidally disrupted outside the inner 1 au from the star.

Two outcomes are possible for these planets. If they accrete gas rapidly, they become brown dwarfs or even low mass stellar companions, just as explained above.

{\em Migration -- formation of planets inward of $\sim 10$~au.} In the opposite case, post collapse planets do not accrete gas. These planets migrate rapidly in but they are not tidally disrupted because they are quite dense to begin with. In the short simulations presented here, the planets migrated to separations of $a \sim 10 - 20$~au and then stalled there. With the initial inner disc boundary set at 10~au this is to be expected: there is little gas material there for the planet migration to continue. However, gas giant planets usually open deep gaps in the protoplanetary discs when they get to $\sim 10$~au \citep{Nayakshin15c} radii because the discs are geometrically thinner there than at the planet's birthplace and these discs are also not self-gravitating. These planets thus could migrate closer in to the star, but at a much slower (type II) migration rate. Where the planets end up inside the inner 10 au region is dependent on the disc evolution on longer time scales not modelled here.

\subsection{Comparison to previous work}\label{sec:previous}

The results presented here are not very unexpected. \cite{OrmelEtal15a,OrmelEtal15} performed simulations of gas accretion onto massive solid cores embedded in gas discs on $\sim 1$~au scales from the host star. Despite the very different mass and length scales, their study has similarities to the work presented here. Their results qualitatively agree with those summarised in \S \ref{sec:main}. For long cooling times the authors found that the atmosphere captured by the planet had a transient character and was not truly bound to the planet. This resulted in a significant reduction of gas accretion rate onto the planet compared with the rate of gas inflow into the Hill sphere of the planet. The same effect is seen in the simulations presented here.

The difficulty of gas clumps surviving as gas giants on wide orbits has also been discussed previously.
\cite{ZhuEtal12a} simulated fragmentation of large scale gas discs in 2D, allowing the discs to be built up self-consistently by the infall of material from larger scales. Rather than varying the dust opacity as was done here, these authors instead used a fixed opacity (equivalent to setting $f_{\rm op}=1$) but varied the deposition rate of gas into the disc. Their main conclusion is very similar to that of the present paper: ".. fast migration, accretion, and tidal destruction of the clumps pose challenges to the scenario of giant planet formation by GI (gravitational instability) in situ...". This is despite the fact that \cite{ZhuEtal12a} study uses very different numerical methods (fixed grid rather than SPH), initial disc set-up, and assumptions. This seems to indicate a broad support to the thesis made in \S \ref{sec:main}: numerical and physical detail do affect which of the three evolutionary paths the clumps take when they evolve out of the desert but the fact that they are not likely to survive in the desert as gas giant planets is independent of those detail.

\cite{NayakshinCha13} and \cite{Stamatellos15} have shown that the outcome of gas clump formation in a massive gas disc may depend strongly on the thermal state of the gas inside the Hill sphere of the clump. Both studies considered the effects of radiative feedback of the clump/planet onto the surrounding gas. The former study focused on pre-collapse gas clumps, whereas the latter looked at post-collapse planets, but both found that radiative preheating of the surrounding disc by the radiation emanating from the planet may significantly reduce the accretion rate onto the planet. The strong radiative preheating regime is effectively same as the long cooling time cases studied here, preventing gas from accreting onto the planet. 

The radiative feedback effect is not included in the simulations presented here in order to not overcomplicate the present study. However results of  \cite{NayakshinCha13} and \cite{Stamatellos15}  add weight to the main conclusions of this paper. Both of these studies showed that when gas accretion is unimportant the planets migrate in rapidly whereas when accretion is fast the planets run away into the brown dwarf regime and stall at wide separations.


\subsection{Observational implications}\label{sec:observations}

Although there are well known massive gas giant planets orbiting their hosts at separations of tens of au, such as HR 8799 system \citep{MaroisEtal08,MaroisEtal10}, statistically there is a significant lack of gas giant planets found at wide separations \citep[e.g.,][]{ViganEtal12,ChauvinEtal15,BowlerEtal15,RiceEtal15}. \cite{BillerEtal13} finds that no more than a few \% of stars host $1-20\mj$ companions with separations in the range $10 - 150$~AU. \cite{GalicherEtal16} find that $\sim 1-2$\% of Solar type stars have a gas giant planet with mass between $0.5\mj$ and $14 \mj$ and separations $20-300$~au.

Results obtained here suggest that we must be careful interpreting these observations. On the one hand, these observations may imply that gravitational disc instability infrequently hatches gas clumps of planetary mass \citep{SW08,KratterEtal10,ForganRice11,ForganRice13}, instead producing {\em only} brown dwarf mass objects. In this case, gravitational disc instability is a rather minor ingredient in planet formation. 

While this interpretation of observational constraints is physically plausible, it appears equally likely that gravitational instability does hatch objects of a few Jupiter mass or less \citep[e.g.,][]{BoleyEtal10} frequently, but that nearly all of these objects evolve away from their birth parameter space (that is, planetary mass and separation of tens to hundreds of au). Migration of planet mass objects is extremely rapid, removing clumps with a few $\mj$ mass from the outer disc and delivering them into the inner $\sim 10-20$ au region in a few thousand years, typically \citep{VB06,BoleyEtal10,MichaelEtal11,BaruteauEtal11,MachidaEtal11,ChaNayakshin11a}. In contrast, protoplanetary discs are known to exist for $\sim 3$ Million years \citep{HaischEtal01}. This implies that, statistically speaking, it is nearly impossible for a gas giant planet to avoid migrating into the inner disc before the disc in removed. 


Objects such as HR 8799 \citep{MaroisEtal08,MaroisEtal10} must have evolved differently than most other gravitationally unstable protoplanetary discs. Perhaps N-body effects of planet-planet interactions slowed down planet migration or even reversed it in this system, or perhaps the parent protoplanetary disc was removed extremely rapidly by a nearby OB star ionising radiation \citep[e.g.,][]{Clarke07}, which is no longer there; swept away by a supernova explosion, as hypothesised for the Solar System, or perturbed by a close star passage, etc.

A calculation of the frequency of the unusual system evolution that could leave a few $\mj$ gas fragments orbiting the host star at tens of au after the parent disc dissipation is beyond the scope of this paper. We may however place some limits on it based on previous work and observations. 

In \cite{Nayakshin16a}, population synthesis of gas clumps born at $\sim 100$~au in massive self-gravitating discs is presented. The study considers only one gas clump per disc and concludes that $\approx 5$\% of gas clumps initially born in the disc survive on distant orbits by the time the disc is dissipated. The rest either migrate into the inner disc or get disrupted at tens of au. Such disruptions are not a null outcome necessarily. If grains grow and sediment rapidly into a massive core \citep{McCreaWilliams65,Boss97,HelledEtal08}, then the disruption leaves behind a massive solid core located beyond $\sim 10$~au initially. The core can then continue to migrate and arrive in the inner disc eventually or may stall at tens of au. The cores stalled at tens of au separations may present an attractive model \citep{Nayakshin16a} for the formation of the suspected surprisingly young planets in the HL Tau protoplanetary disc \citep{BroganEtal15,PinteEtal16,JinEtal16,PicognaK15,DipierroEtal16a,RosottiEtal16}.

Since the study of \cite{Nayakshin16a} does not take into account gas accretion onto the fragments, the $\sim 5$\% frequency of gas clump survival at wide separations may be considered as the upper limit. If this estimate is correct then the observed $\sim 1$\% frequency of gas giant planets in wide orbits requires that at least $\sim 20$\% of young protoplanetary discs fragmented on planetary mass objects at wide separations.

A lower limit on the survival probability of gas giant planets may be obtained by assuming that all FU Ori outbursts result from tidal disruptions of gas clumps created by gravitational instability in the outer disc after they migrated into the inner disc \citep{VB05,VB06,VB10,BoleyEtal10,NayakshinLodato12}. Since statistical arguments suggest that there is a dozen FU Ori events per protostar \citep{HK96}, we require $\sim 10$ planet mass clumps per star on average. Under these assumptions, the observed frequency of gas giants imaged beyond 10 au \citep{GalicherEtal16} then suggests that only $\sim 1$ in a thousand of such clumps survive on distant orbits to the present day. In this picture each star makes a dozen planet mass gas clumps at wide separations.

Summarising, it is crucially important to understand whether gravitationally fragmenting discs never make planet mass clumps or they make many but 99.9\% of them are destroyed by migration, gas accretion or tidal disruptions. To resolve the issue, we need more detailed theoretical models and also more observations of stellar companions in the massive planet and the brown dwarf regimes. If gravitational disc instability only makes brown dwarfs then that population is likely to be completely different from the population of massive planets in terms of mass versus orbital separation distributions, host metallicity correlations, bulk metallicity content, etc. If, on the other hand, gravitational disc fragmentation produces planetary mass clumps and they then evolve into the three groups of objects as shown in fig. \ref{fig:MoneyPlot}, then planets and brown dwarfs must be related to one another and thus share some characteristics. Note that in this case $\sim 1\mj$ gas giants may still be different in properties from brown dwarfs but the transition in properties between the two populations must be continuos and explainable by some robust physics. Future work will undoubtedly establish which of the two paradigms is correct and will therefore greatly contribute to testing the planet formation scenarios.

\section{Conclusions}\label{sec:conclusions}

Simulations presented in this paper show that gas giant mass clumps born by the gravitational instability in the outer cold disc are very unlikely to remain there after the disc dissipation. These clumps are extremely efficiently removed from that region by a trio of processes: (1) inward planet migration; (2) runaway accretion making them brown dwarfs or even low mass stars; (3) tidal disruption downsizing them into sub-giant planetary domain. This implies that the observed fraction of just $\sim 1$\% of gas giant planets orbiting their hosts outside $\sim 10$ au region may be just a tip of an iceberg of the initially numerous population of planetary mass clumps formed by gravitational instability in that region. Further theoretical modelling and observations of the massive planet -- brown dwarf transition are needed to shed light on the actual frequency of disc fragmentation onto clumps of gas giant planet mass at $\sim 100$~au.


\begin{thebibliography}{}
\makeatletter
\relax
\def\mn@urlcharsother{\let\do\@makeother \do\$\do\&\do\#\do\^\do\_\do\%\do\~}
\def\mn@doi{\begingroup\mn@urlcharsother \@ifnextchar [ {\mn@doi@}
  {\mn@doi@[]}}
\def\mn@doi@[#1]#2{\def\@tempa{#1}\ifx\@tempa\@empty \href
  {http://dx.doi.org/#2} {doi:#2}\else \href {http://dx.doi.org/#2} {#1}\fi
  \endgroup}
\def\mn@eprint#1#2{\mn@eprint@#1:#2::\@nil}
\def\mn@eprint@arXiv#1{\href {http://arxiv.org/abs/#1} {{\tt arXiv:#1}}}
\def\mn@eprint@dblp#1{\href {http://dblp.uni-trier.de/rec/bibtex/#1.xml}
  {dblp:#1}}
\def\mn@eprint@#1:#2:#3:#4\@nil{\def\@tempa {#1}\def\@tempb {#2}\def\@tempc
  {#3}\ifx \@tempc \@empty \let \@tempc \@tempb \let \@tempb \@tempa \fi \ifx
  \@tempb \@empty \def\@tempb {arXiv}\fi \@ifundefined
  {mn@eprint@\@tempb}{\@tempb:\@tempc}{\expandafter \expandafter \csname
  mn@eprint@\@tempb\endcsname \expandafter{\@tempc}}}

\bibitem[\protect\citeauthoryear{{Baruteau}, {Meru}  \&
  {Paardekooper}}{{Baruteau} et~al.}{2011}]{BaruteauEtal11}
{Baruteau} C.,  {Meru} F.,   {Paardekooper} S.-J.,  2011, \mn@doi [\mnras]
  {10.1111/j.1365-2966.2011.19172.x}, \href
  {http://adsabs.harvard.edu/abs/2011MNRAS.416.1971B} {416, 1971}

\bibitem[\protect\citeauthoryear{{Bate}, {Bonnell}  \& {Price}}{{Bate}
  et~al.}{1995}]{Bate95}
{Bate} M.~R.,  {Bonnell} I.~A.,   {Price} N.~M.,  1995, \mnras, \href
  {http://adsabs.harvard.edu/cgi-bin/nph-bib_query?bibcode=1995MNRAS.277..362B&db_key=AST}
  {277, 362}

\bibitem[\protect\citeauthoryear{{Biller} et~al.,}{{Biller}
  et~al.}{2013}]{BillerEtal13}
{Biller} B.~A.,  et~al., 2013, \mn@doi [\apj] {10.1088/0004-637X/777/2/160},
  \href {http://adsabs.harvard.edu/abs/2013ApJ...777..160B} {777, 160}

\bibitem[\protect\citeauthoryear{{Bodenheimer}}{{Bodenheimer}}{1974}]{Bodenheimer74}
{Bodenheimer} P.,  1974, \mn@doi [Icarus] {10.1016/0019-1035(74)90050-5}, \href
  {http://ukads.nottingham.ac.uk/abs/1974Icar...23..319B} {23, 319}

\bibitem[\protect\citeauthoryear{{Boley}}{{Boley}}{2009}]{Boley09}
{Boley} A.~C.,  2009, \mn@doi [\apjl] {10.1088/0004-637X/695/1/L53}, \href
  {http://adsabs.harvard.edu/abs/2009ApJ...695L..53B} {695, L53}

\bibitem[\protect\citeauthoryear{{Boley}, {Hayfield}, {Mayer}  \&
  {Durisen}}{{Boley} et~al.}{2010}]{BoleyEtal10}
{Boley} A.~C.,  {Hayfield} T.,  {Mayer} L.,   {Durisen} R.~H.,  2010, \mn@doi
  [Icarus] {10.1016/j.icarus.2010.01.015}, \href
  {http://ukads.nottingham.ac.uk/abs/2010Icar..207..509B} {207, 509}

\bibitem[\protect\citeauthoryear{{Boss}}{{Boss}}{1997}]{Boss97}
{Boss} A.~P.,  1997, \mn@doi [Science] {10.1126/science.276.5320.1836}, \href
  {http://ukads.nottingham.ac.uk/abs/1997Sci...276.1836B} {276, 1836}

\bibitem[\protect\citeauthoryear{{Bowler}, {Liu}, {Shkolnik}  \&
  {Tamura}}{{Bowler} et~al.}{2015}]{BowlerEtal15}
{Bowler} B.~P.,  {Liu} M.~C.,  {Shkolnik} E.~L.,   {Tamura} M.,  2015, \mn@doi
  [\apjs] {10.1088/0067-0049/216/1/7}, \href
  {http://adsabs.harvard.edu/abs/2015ApJS..216....7B} {216, 7}

\bibitem[\protect\citeauthoryear{{Brogan} et~al.,}{{Brogan}
  et~al.}{2015}]{BroganEtal15}
{Brogan} C.~L.,  et~al., 2015, preprint, \href
  {http://adsabs.harvard.edu/abs/2015arXiv150302649P} {} (\mn@eprint {arXiv}
  {1503.02649})

\bibitem[\protect\citeauthoryear{{Burrows}, {Hubbard}, {Lunine}  \&
  {Liebert}}{{Burrows} et~al.}{2001}]{BurrowsEtal01}
{Burrows} A.,  {Hubbard} W.~B.,  {Lunine} J.~I.,   {Liebert} J.,  2001, \mn@doi
  [Reviews of Modern Physics] {10.1103/RevModPhys.73.719}, \href
  {http://adsabs.harvard.edu/abs/2001RvMP...73..719B} {73, 719}

\bibitem[\protect\citeauthoryear{{Cha} \& {Nayakshin}}{{Cha} \&
  {Nayakshin}}{2011}]{ChaNayakshin11a}
{Cha} S.-H.,  {Nayakshin} S.,  2011, \mn@doi [\mnras]
  {10.1111/j.1365-2966.2011.18953.x}, \href
  {http://adsabs.harvard.edu/abs/2011MNRAS.415.3319C} {415, 3319}

\bibitem[\protect\citeauthoryear{{Chauvin} et~al.,}{{Chauvin}
  et~al.}{2015}]{ChauvinEtal15}
{Chauvin} G.,  et~al., 2015, \mn@doi [\aap] {10.1051/0004-6361/201423564},
  \href {http://adsabs.harvard.edu/abs/2015A%26A...573A.127C} {573, A127}

\bibitem[\protect\citeauthoryear{{Clarke}}{{Clarke}}{2007}]{Clarke07}
{Clarke} C.~J.,  2007, \mn@doi [\mnras] {10.1111/j.1365-2966.2007.11547.x},
  \href {http://adsabs.harvard.edu/abs/2007MNRAS.376.1350C} {376, 1350}

\bibitem[\protect\citeauthoryear{{Cossins}, {Lodato}  \& {Clarke}}{{Cossins}
  et~al.}{2009}]{CossinsEtal09}
{Cossins} P.,  {Lodato} G.,   {Clarke} C.~J.,  2009, \mn@doi [\mnras]
  {10.1111/j.1365-2966.2008.14275.x}, \href
  {http://adsabs.harvard.edu/abs/2009MNRAS.393.1157C} {393, 1157}

\bibitem[\protect\citeauthoryear{{Crida}, {Morbidelli}  \& {Masset}}{{Crida}
  et~al.}{2006}]{CridaEtal06}
{Crida} A.,  {Morbidelli} A.,   {Masset} F.,  2006, \mn@doi [\icarus]
  {10.1016/j.icarus.2005.10.007}, \href
  {http://adsabs.harvard.edu/abs/2006Icar..181..587C} {181, 587}

\bibitem[\protect\citeauthoryear{{Cuadra}, {Nayakshin}, {Springel}  \& {Di
  Matteo}}{{Cuadra} et~al.}{2006}]{Cuadra06}
{Cuadra} J.,  {Nayakshin} S.,  {Springel} V.,   {Di Matteo} T.,  2006, \mn@doi
  [\mnras] {10.1111/j.1365-2966.2005.09837.x}, \href
  {http://ukads.nottingham.ac.uk/cgi-bin/nph-bib_query?bibcode=2006MNRAS.366..358C&db_key=AST}
  {366, 358}

\bibitem[\protect\citeauthoryear{{Dipierro}, {Laibe}, {Price}  \&
  {Lodato}}{{Dipierro} et~al.}{2016}]{DipierroEtal16a}
{Dipierro} G.,  {Laibe} G.,  {Price} D.~J.,   {Lodato} G.,  2016, \mn@doi
  [\mnras] {10.1093/mnrasl/slw032}, \href
  {http://adsabs.harvard.edu/abs/2016MNRAS.tmpL..16D} {}

\bibitem[\protect\citeauthoryear{{Dullemond} \& {Dominik}}{{Dullemond} \&
  {Dominik}}{2005}]{DD05}
{Dullemond} C.~P.,  {Dominik} C.,  2005, \mn@doi [\aap]
  {10.1051/0004-6361:20042080}, \href
  {http://ukads.nottingham.ac.uk/abs/2005A%26A...434..971D} {434, 971}

\bibitem[\protect\citeauthoryear{{Durisen}, {Boss}, {Mayer}, {Nelson}, {Quinn}
  \& {Rice}}{{Durisen} et~al.}{2007}]{DurisenEtal07}
{Durisen} R.~H.,  {Boss} A.~P.,  {Mayer} L.,  {Nelson} A.~F.,  {Quinn} T.,
  {Rice} W.~K.~M.,  2007, Protostars and Planets V, \href
  {http://adsabs.harvard.edu/abs/2007prpl.conf..607D} {pp 607--622}

\bibitem[\protect\citeauthoryear{{Forgan} \& {Rice}}{{Forgan} \&
  {Rice}}{2011}]{ForganRice11}
{Forgan} D.,  {Rice} K.,  2011, \mn@doi [\mnras]
  {10.1111/j.1365-2966.2011.19380.x}, \href
  {http://adsabs.harvard.edu/abs/2011MNRAS.417.1928F} {417, 1928}

\bibitem[\protect\citeauthoryear{{Forgan} \& {Rice}}{{Forgan} \&
  {Rice}}{2013a}]{ForganRice13}
{Forgan} D.,  {Rice} K.,  2013a, \mn@doi [\mnras] {10.1093/mnras/stt032}, \href
  {http://adsabs.harvard.edu/abs/2013MNRAS.430.2082F} {430, 2082}

\bibitem[\protect\citeauthoryear{{Forgan} \& {Rice}}{{Forgan} \&
  {Rice}}{2013b}]{ForganRice13b}
{Forgan} D.,  {Rice} K.,  2013b, \mn@doi [\mnras] {10.1093/mnras/stt672}, \href
  {http://adsabs.harvard.edu/abs/2013MNRAS.432.3168F} {432, 3168}

\bibitem[\protect\citeauthoryear{{Galicher} et~al.,}{{Galicher}
  et~al.}{2016}]{GalicherEtal16}
{Galicher} R.,  et~al., 2016, preprint, \href
  {http://adsabs.harvard.edu/abs/2016arXiv160708239G} {} (\mn@eprint {arXiv}
  {1607.08239})

\bibitem[\protect\citeauthoryear{{Galvagni}, {Hayfield}, {Boley}, {Mayer},
  {Ro{\v s}kar}  \& {Saha}}{{Galvagni} et~al.}{2012}]{GalvagniEtal12}
{Galvagni} M.,  {Hayfield} T.,  {Boley} A.,  {Mayer} L.,  {Ro{\v s}kar} R.,
  {Saha} P.,  2012, \mn@doi [\mnras] {10.1111/j.1365-2966.2012.22096.x}, \href
  {http://adsabs.harvard.edu/abs/2012MNRAS.427.1725G} {427, 1725}

\bibitem[\protect\citeauthoryear{{Gammie}}{{Gammie}}{2001}]{Gammie01}
{Gammie} C.~F.,  2001, \apj, \href
  {http://cdsads.u-strasbg.fr/cgi-bin/nph-bib_query?bibcode=2001ApJ...553..174G&amp;db_key=AST}
  {553, 174}

\bibitem[\protect\citeauthoryear{{Goldreich} \& {Ward}}{{Goldreich} \&
  {Ward}}{1973}]{GoldreichWard73}
{Goldreich} P.,  {Ward} W.~R.,  1973, \mn@doi [\apj] {10.1086/152291}, \href
  {http://adsabs.harvard.edu/abs/1973ApJ...183.1051G} {183, 1051}

\bibitem[\protect\citeauthoryear{{Haisch}, {Lada}  \& {Lada}}{{Haisch}
  et~al.}{2001}]{HaischEtal01}
{Haisch} Jr. K.~E.,  {Lada} E.~A.,   {Lada} C.~J.,  2001, \mn@doi [\apjl]
  {10.1086/320685}, \href {http://adsabs.harvard.edu/abs/2001ApJ...553L.153H}
  {553, L153}

\bibitem[\protect\citeauthoryear{{Hartmann} \& {Kenyon}}{{Hartmann} \&
  {Kenyon}}{1996}]{HK96}
{Hartmann} L.,  {Kenyon} S.~J.,  1996, \mn@doi [\araa]
  {10.1146/annurev.astro.34.1.207}, \href
  {http://adsabs.harvard.edu/abs/1996ARA%26A..34..207H} {34, 207}

\bibitem[\protect\citeauthoryear{{Haworth} et~al.,}{{Haworth}
  et~al.}{2016}]{HaworthEtal16}
{Haworth} T.~J.,  et~al., 2016, preprint, \href
  {http://adsabs.harvard.edu/abs/2016arXiv160801315H} {} (\mn@eprint {arXiv}
  {1608.01315})

\bibitem[\protect\citeauthoryear{{Helled} \& {Bodenheimer}}{{Helled} \&
  {Bodenheimer}}{2011}]{HB11}
{Helled} R.,  {Bodenheimer} P.,  2011, \mn@doi [\icarus]
  {10.1016/j.icarus.2010.09.024}, \href
  {http://adsabs.harvard.edu/abs/2011Icar..211..939H} {211, 939}

\bibitem[\protect\citeauthoryear{{Helled} \& {Schubert}}{{Helled} \&
  {Schubert}}{2008}]{HS08}
{Helled} R.,  {Schubert} G.,  2008, \mn@doi [Icarus]
  {10.1016/j.icarus.2008.08.002}, \href
  {http://ukads.nottingham.ac.uk/abs/2008Icar..198..156H} {198, 156}

\bibitem[\protect\citeauthoryear{{Helled}, {Podolak}  \& {Kovetz}}{{Helled}
  et~al.}{2008}]{HelledEtal08}
{Helled} R.,  {Podolak} M.,   {Kovetz} A.,  2008, \mn@doi [Icarus]
  {10.1016/j.icarus.2008.01.007}, \href
  {http://ukads.nottingham.ac.uk/abs/2008Icar..195..863H} {195, 863}

\bibitem[\protect\citeauthoryear{{Helled} et~al.,}{{Helled}
  et~al.}{2014}]{HelledEtal13a}
{Helled} R.,  et~al., 2014, \mn@doi [Protostars and Planets VI, University of
  Arizona Press, Tucson] {10.2458/azu_uapress_9780816531240-ch028}, \href
  {http://adsabs.harvard.edu/abs/2014prpl.conf..643H} {pp 643--665}

\bibitem[\protect\citeauthoryear{{Inutsuka}, {Machida}  \&
  {Matsumoto}}{{Inutsuka} et~al.}{2010}]{InutsukaEtal09}
{Inutsuka} S.,  {Machida} M.~N.,   {Matsumoto} T.,  2010, \mn@doi [\apjl]
  {10.1088/2041-8205/718/2/L58}, \href
  {http://adsabs.harvard.edu/abs/2010ApJ...718L..58I} {718, L58}

\bibitem[\protect\citeauthoryear{{Jin}, {Li}, {Isella}, {Li}  \& {Ji}}{{Jin}
  et~al.}{2016}]{JinEtal16}
{Jin} S.,  {Li} S.,  {Isella} A.,  {Li} H.,   {Ji} J.,  2016, preprint, \href
  {http://adsabs.harvard.edu/abs/2016arXiv160100358J} {} (\mn@eprint {arXiv}
  {1601.00358})

\bibitem[\protect\citeauthoryear{{Johnson} \& {Gammie}}{{Johnson} \&
  {Gammie}}{2003}]{Johnson03}
{Johnson} B.~M.,  {Gammie} C.~F.,  2003, \apj, \href
  {http://cdsads.u-strasbg.fr/cgi-bin/nph-bib_query?bibcode=2003ApJ...597..131J&amp;db_key=AST}
  {597, 131}

\bibitem[\protect\citeauthoryear{{Kratter} \& {Lodato}}{{Kratter} \&
  {Lodato}}{2016}]{KratterL16}
{Kratter} K.~M.,  {Lodato} G.,  2016, preprint, \href
  {http://adsabs.harvard.edu/abs/2016arXiv160301280K} {} (\mn@eprint {arXiv}
  {1603.01280})

\bibitem[\protect\citeauthoryear{{Kratter}, {Murray-Clay}  \&
  {Youdin}}{{Kratter} et~al.}{2010}]{KratterEtal10}
{Kratter} K.~M.,  {Murray-Clay} R.~A.,   {Youdin} A.~N.,  2010, \mn@doi [\apj]
  {10.1088/0004-637X/710/2/1375}, \href
  {http://adsabs.harvard.edu/abs/2010ApJ...710.1375K} {710, 1375}

\bibitem[\protect\citeauthoryear{{Kuiper}}{{Kuiper}}{1951}]{Kuiper51b}
{Kuiper} G.~P.,  1951, \mn@doi [Proceedings of the National Academy of Science]
  {10.1073/pnas.37.1.1}, \href
  {http://adsabs.harvard.edu/abs/1951PNAS...37....1K} {37, 1}

\bibitem[\protect\citeauthoryear{{Kumar}}{{Kumar}}{1972}]{Kumar72}
{Kumar} S.~S.,  1972, \mn@doi [\apss] {10.1007/BF00643091}, \href
  {http://adsabs.harvard.edu/abs/1972Ap%26SS..16...52K} {16, 52}

\bibitem[\protect\citeauthoryear{{Larson}}{{Larson}}{1969}]{Larson69}
{Larson} R.~B.,  1969, \mnras, \href
  {http://ukads.nottingham.ac.uk/cgi-bin/nph-bib_query?bibcode=1969MNRAS.145..271L&db_key=AST}
  {145, 271}

\bibitem[\protect\citeauthoryear{{Laughlin} \& {Bodenheimer}}{{Laughlin} \&
  {Bodenheimer}}{1994}]{LaughlinBodenheimer94}
{Laughlin} G.,  {Bodenheimer} P.,  1994, \mn@doi [\apj] {10.1086/174909}, \href
  {http://adsabs.harvard.edu/abs/1994ApJ...436..335L} {436, 335}

\bibitem[\protect\citeauthoryear{{Lin} \& {Kratter}}{{Lin} \&
  {Kratter}}{2016}]{LinKratter16}
{Lin} M.-K.,  {Kratter} K.~M.,  2016, \mn@doi [\apj]
  {10.3847/0004-637X/824/2/91}, \href
  {http://adsabs.harvard.edu/abs/2016ApJ...824...91L} {824, 91}

\bibitem[\protect\citeauthoryear{{Lissauer}}{{Lissauer}}{1987}]{Lissauer87}
{Lissauer} J.~J.,  1987, Icarus, \href
  {http://esoads.eso.org/cgi-bin/nph-bib_query?bibcode=1987Icar...69..249L&amp;db_key=AST}
  {69, 249}

\bibitem[\protect\citeauthoryear{{Machida}, {Inutsuka}  \&
  {Matsumoto}}{{Machida} et~al.}{2011}]{MachidaEtal11}
{Machida} M.~N.,  {Inutsuka} S.-i.,   {Matsumoto} T.,  2011, \mn@doi [\apj]
  {10.1088/0004-637X/729/1/42}, \href
  {http://adsabs.harvard.edu/abs/2011ApJ...729...42M} {729, 42}

\bibitem[\protect\citeauthoryear{{Malik}, {Meru}, {Mayer}  \& {Meyer}}{{Malik}
  et~al.}{2015}]{MalikEtal15}
{Malik} M.,  {Meru} F.,  {Mayer} L.,   {Meyer} M.,  2015, \mn@doi [\apj]
  {10.1088/0004-637X/802/1/56}, \href
  {http://adsabs.harvard.edu/abs/2015ApJ...802...56M} {802, 56}

\bibitem[\protect\citeauthoryear{{Marois}, {Macintosh}, {Barman}, {Zuckerman},
  {Song}, {Patience}, {Lafreni{\`e}re}  \& {Doyon}}{{Marois}
  et~al.}{2008}]{MaroisEtal08}
{Marois} C.,  {Macintosh} B.,  {Barman} T.,  {Zuckerman} B.,  {Song} I.,
  {Patience} J.,  {Lafreni{\`e}re} D.,   {Doyon} R.,  2008, \mn@doi [Science]
  {10.1126/science.1166585}, \href
  {http://adsabs.harvard.edu/abs/2008Sci...322.1348M} {322, 1348}

\bibitem[\protect\citeauthoryear{{Marois}, {Zuckerman}, {Konopacky},
  {Macintosh}  \& {Barman}}{{Marois} et~al.}{2010}]{MaroisEtal10}
{Marois} C.,  {Zuckerman} B.,  {Konopacky} Q.~M.,  {Macintosh} B.,   {Barman}
  T.,  2010, \mn@doi [\nat] {10.1038/nature09684}, \href
  {http://adsabs.harvard.edu/abs/2010Natur.468.1080M} {468, 1080}

\bibitem[\protect\citeauthoryear{{Mayer}, {Quinn}, {Wadsley}  \&
  {Stadel}}{{Mayer} et~al.}{2004}]{MayerEtal04}
{Mayer} L.,  {Quinn} T.,  {Wadsley} J.,   {Stadel} J.,  2004, \mn@doi [\apj]
  {10.1086/421288}, \href {http://adsabs.harvard.edu/abs/2004ApJ...609.1045M}
  {609, 1045}

\bibitem[\protect\citeauthoryear{{Mayor} et~al.,}{{Mayor}
  et~al.}{2011}]{MayorEtal11}
{Mayor} M.,  et~al., 2011, ArXiv e-prints (astro-ph 1109.2497), \href
  {http://adsabs.harvard.edu/abs/2011arXiv1109.2497M} {}

\bibitem[\protect\citeauthoryear{{McCrea} \& {Williams}}{{McCrea} \&
  {Williams}}{1965}]{McCreaWilliams65}
{McCrea} W.~H.,  {Williams} I.~P.,  1965, Royal Society of London Proceedings
  Series A, \href {http://adsabs.harvard.edu/abs/1965RSPSA.287..143M} {287,
  143}

\bibitem[\protect\citeauthoryear{{Meru} \& {Bate}}{{Meru} \&
  {Bate}}{2011}]{MeruBate11a}
{Meru} F.,  {Bate} M.~R.,  2011, \mn@doi [\mnras]
  {10.1111/j.1745-3933.2010.00978.x}, \href
  {http://adsabs.harvard.edu/abs/2011MNRAS.411L...1M} {411, L1}

\bibitem[\protect\citeauthoryear{{Meru} \& {Bate}}{{Meru} \&
  {Bate}}{2012}]{MeruBate12}
{Meru} F.,  {Bate} M.~R.,  2012, \mn@doi [\mnras]
  {10.1111/j.1365-2966.2012.22035.x}, \href
  {http://adsabs.harvard.edu/abs/2012MNRAS.427.2022M} {427, 2022}

\bibitem[\protect\citeauthoryear{{Michael}, {Durisen}  \& {Boley}}{{Michael}
  et~al.}{2011}]{MichaelEtal11}
{Michael} S.,  {Durisen} R.~H.,   {Boley} A.~C.,  2011, \mn@doi [\apjl]
  {10.1088/2041-8205/737/2/L42}, \href
  {http://adsabs.harvard.edu/abs/2011ApJ...737L..42M} {737, L42+}

\bibitem[\protect\citeauthoryear{{Michael}, {Steiman-Cameron}, {Durisen}  \&
  {Boley}}{{Michael} et~al.}{2012}]{MichaelEtal12}
{Michael} S.,  {Steiman-Cameron} T.~Y.,  {Durisen} R.~H.,   {Boley} A.~C.,
  2012, \mn@doi [\apj] {10.1088/0004-637X/746/1/98}, \href
  {http://adsabs.harvard.edu/abs/2012ApJ...746...98M} {746, 98}

\bibitem[\protect\citeauthoryear{{Nayakshin}}{{Nayakshin}}{2010a}]{Nayakshin10c}
{Nayakshin} S.,  2010a, \mn@doi [\mnras] {10.1111/j.1745-3933.2010.00923.x},
  \href {http://adsabs.harvard.edu/abs/2010MNRAS.408L..36N} {408, L36}

\bibitem[\protect\citeauthoryear{{Nayakshin}}{{Nayakshin}}{2010b}]{Nayakshin10a}
{Nayakshin} S.,  2010b, \mn@doi [\mnras] {10.1111/j.1365-2966.2010.17289.x},
  \href {http://adsabs.harvard.edu/abs/2010MNRAS.408.2381N} {408, 2381}

\bibitem[\protect\citeauthoryear{{Nayakshin}}{{Nayakshin}}{2011}]{Nayakshin10b}
{Nayakshin} S.,  2011, \mn@doi [\mnras] {10.1111/j.1365-2966.2011.18230.x},
  \href {http://adsabs.harvard.edu/abs/2011MNRAS.413.1462N} {413, 1462}

\bibitem[\protect\citeauthoryear{{Nayakshin}}{{Nayakshin}}{2015a}]{Nayakshin15d}
{Nayakshin} S.,  2015a, ArXiv e-prints (arXiv: 1502.07585), \href
  {http://adsabs.harvard.edu/abs/2014arXiv1411.5264N} {}

\bibitem[\protect\citeauthoryear{{Nayakshin}}{{Nayakshin}}{2015b}]{Nayakshin15a}
{Nayakshin} S.,  2015b, \mn@doi [\mnras] {10.1093/mnras/stu2074}, \href
  {http://adsabs.harvard.edu/abs/2015MNRAS.446..459N} {446, 459}

\bibitem[\protect\citeauthoryear{{Nayakshin}}{{Nayakshin}}{2015c}]{Nayakshin15c}
{Nayakshin} S.,  2015c, \mn@doi [\mnras] {10.1093/mnras/stv1915}, \href
  {http://adsabs.harvard.edu/abs/2015MNRAS.454...64N} {454, 64}

\bibitem[\protect\citeauthoryear{{Nayakshin}}{{Nayakshin}}{2016a}]{Nayakshin_Review}
{Nayakshin} S.,  2016a, preprint, \href
  {http://adsabs.harvard.edu/abs/2016arXiv160907503N} {} (\mn@eprint {arXiv}
  {1609.07503})

\bibitem[\protect\citeauthoryear{{Nayakshin}}{{Nayakshin}}{2016b}]{Nayakshin16a}
{Nayakshin} S.,  2016b, \mn@doi [\mnras] {10.1093/mnras/stw1404}, \href
  {http://adsabs.harvard.edu/abs/2016MNRAS.461.3194N} {461, 3194}

\bibitem[\protect\citeauthoryear{{Nayakshin} \& {Cha}}{{Nayakshin} \&
  {Cha}}{2013}]{NayakshinCha13}
{Nayakshin} S.,  {Cha} S.-H.,  2013, \mn@doi [\mnras] {10.1093/mnras/stt1426},
  \href {http://adsabs.harvard.edu/abs/2013MNRAS.435.2099N} {435, 2099}

\bibitem[\protect\citeauthoryear{{Nayakshin} \& {Lodato}}{{Nayakshin} \&
  {Lodato}}{2012}]{NayakshinLodato12}
{Nayakshin} S.,  {Lodato} G.,  2012, \mn@doi [\mnras]
  {10.1111/j.1365-2966.2012.21612.x}, \href
  {http://adsabs.harvard.edu/abs/2012MNRAS.426...70N} {426, 70}

\bibitem[\protect\citeauthoryear{{Ormel}, {Kuiper}  \& {Shi}}{{Ormel}
  et~al.}{2015a}]{OrmelEtal15a}
{Ormel} C.~W.,  {Kuiper} R.,   {Shi} J.-M.,  2015a, \mn@doi [\mnras]
  {10.1093/mnras/stu2101}, \href
  {http://adsabs.harvard.edu/abs/2015MNRAS.446.1026O} {446, 1026}

\bibitem[\protect\citeauthoryear{{Ormel}, {Shi}  \& {Kuiper}}{{Ormel}
  et~al.}{2015b}]{OrmelEtal15}
{Ormel} C.~W.,  {Shi} J.-M.,   {Kuiper} R.,  2015b, \mn@doi [\mnras]
  {10.1093/mnras/stu2704}, \href
  {http://adsabs.harvard.edu/abs/2015MNRAS.447.3512O} {447, 3512}

\bibitem[\protect\citeauthoryear{{Paardekooper}}{{Paardekooper}}{2012}]{Paardekooper12a}
{Paardekooper} S.-J.,  2012, \mn@doi [\mnras]
  {10.1111/j.1365-2966.2012.20553.x}, \href
  {http://adsabs.harvard.edu/abs/2012MNRAS.421.3286P} {421, 3286}

\bibitem[\protect\citeauthoryear{{Picogna} \& {Kley}}{{Picogna} \&
  {Kley}}{2015}]{PicognaK15}
{Picogna} G.,  {Kley} W.,  2015, \mn@doi [\aap] {10.1051/0004-6361/201526921},
  \href {http://adsabs.harvard.edu/abs/2015A%26A...584A.110P} {584, A110}

\bibitem[\protect\citeauthoryear{{Pinte}, {Dent}, {M{\'e}nard}, {Hales},
  {Hill}, {Cortes}  \& {de Gregorio-Monsalvo}}{{Pinte}
  et~al.}{2016}]{PinteEtal16}
{Pinte} C.,  {Dent} W.~R.~F.,  {M{\'e}nard} F.,  {Hales} A.,  {Hill} T.,
  {Cortes} P.,   {de Gregorio-Monsalvo} I.,  2016, \mn@doi [\apj]
  {10.3847/0004-637X/816/1/25}, \href
  {http://adsabs.harvard.edu/abs/2016ApJ...816...25P} {816, 25}

\bibitem[\protect\citeauthoryear{{Rafikov}}{{Rafikov}}{2005}]{Rafikov05}
{Rafikov} R.~R.,  2005, \mn@doi [\apjl] {10.1086/428899}, \href
  {http://ukads.nottingham.ac.uk/cgi-bin/nph-bib_query?bibcode=2005ApJ...621L..69R&db_key=AST}
  {621, L69}

\bibitem[\protect\citeauthoryear{{Rice}, {Lodato}  \& {Armitage}}{{Rice}
  et~al.}{2005}]{Rice05}
{Rice} W.~K.~M.,  {Lodato} G.,   {Armitage} P.~J.,  2005, \mn@doi [\mnras]
  {10.1111/j.1745-3933.2005.00105.x}, \href
  {http://ukads.nottingham.ac.uk/cgi-bin/nph-bib_query?bibcode=2005MNRAS.364L..56R&db_key=AST}
  {364, L56}

\bibitem[\protect\citeauthoryear{{Rice}, {Forgan}  \& {Armitage}}{{Rice}
  et~al.}{2012}]{RiceEtal12}
{Rice} W.~K.~M.,  {Forgan} D.~H.,   {Armitage} P.~J.,  2012, \mn@doi [\mnras]
  {10.1111/j.1365-2966.2011.20153.x}, \href
  {http://adsabs.harvard.edu/abs/2012MNRAS.420.1640R} {420, 1640}

\bibitem[\protect\citeauthoryear{{Rice}, {Paardekooper}, {Forgan}  \&
  {Armitage}}{{Rice} et~al.}{2014}]{RiceEtal14a}
{Rice} W.~K.~M.,  {Paardekooper} S.-J.,  {Forgan} D.~H.,   {Armitage} P.~J.,
  2014, \mn@doi [\mnras] {10.1093/mnras/stt2297}, \href
  {http://adsabs.harvard.edu/abs/2014MNRAS.tmp....1R} {}

\bibitem[\protect\citeauthoryear{{Rice}, {Lopez}, {Forgan}  \& {Biller}}{{Rice}
  et~al.}{2015}]{RiceEtal15}
{Rice} K.,  {Lopez} E.,  {Forgan} D.,   {Biller} B.,  2015, preprint, \href
  {http://adsabs.harvard.edu/abs/2015arXiv150806528R} {} (\mn@eprint {arXiv}
  {1508.06528})

\bibitem[\protect\citeauthoryear{{Rogers} \& {Wadsley}}{{Rogers} \&
  {Wadsley}}{2012}]{RogersWadsley12}
{Rogers} P.~D.,  {Wadsley} J.,  2012, \mn@doi [\mnras]
  {10.1111/j.1365-2966.2012.21014.x}, \href
  {http://adsabs.harvard.edu/abs/2012MNRAS.423.1896R} {423, 1896}

\bibitem[\protect\citeauthoryear{{Rosotti}, {Juhasz}, {Booth}  \&
  {Clarke}}{{Rosotti} et~al.}{2016}]{RosottiEtal16}
{Rosotti} G.~P.,  {Juhasz} A.,  {Booth} R.~A.,   {Clarke} C.~J.,  2016,
  preprint, \href {http://adsabs.harvard.edu/abs/2016arXiv160302141R} {}
  (\mn@eprint {arXiv} {1603.02141})

\bibitem[\protect\citeauthoryear{{Semenov}, {Henning}, {Helling}, {Ilgner}  \&
  {Sedlmayr}}{{Semenov} et~al.}{2003}]{SemenovEtal03}
{Semenov} D.,  {Henning} T.,  {Helling} C.,  {Ilgner} M.,   {Sedlmayr} E.,
  2003, \mn@doi [\aap] {10.1051/0004-6361:20031279}, \href
  {http://adsabs.harvard.edu/abs/2003A%26A...410..611S} {410, 611}

\bibitem[\protect\citeauthoryear{{Shakura} \& {Sunyaev}}{{Shakura} \&
  {Sunyaev}}{1973}]{Shakura73}
{Shakura} N.~I.,  {Sunyaev} R.~A.,  1973, \aap, \href
  {http://cdsads.u-strasbg.fr/cgi-bin/nph-bib_query?bibcode=1973A%26A....24..337S&db_key=AST}
  {24, 337}

\bibitem[\protect\citeauthoryear{{Stamatellos}}{{Stamatellos}}{2015}]{Stamatellos15}
{Stamatellos} D.,  2015, \mn@doi [\apjl] {10.1088/2041-8205/810/1/L11}, \href
  {http://adsabs.harvard.edu/abs/2015ApJ...810L..11S} {810, L11}

\bibitem[\protect\citeauthoryear{{Stamatellos} \& {Whitworth}}{{Stamatellos} \&
  {Whitworth}}{2008}]{SW08}
{Stamatellos} D.,  {Whitworth} A.~P.,  2008, \mn@doi [\aap]
  {10.1051/0004-6361:20078628}, \href
  {http://ukads.nottingham.ac.uk/abs/2008A%26A...480..879S} {480, 879}

\bibitem[\protect\citeauthoryear{{Toomre}}{{Toomre}}{1964}]{Toomre64}
{Toomre} A.,  1964, \apj, \href
  {http://esoads.eso.org/cgi-bin/nph-bib_query?bibcode=1964ApJ...139.1217T&amp;db_key=AST}
  {139, 1217}

\bibitem[\protect\citeauthoryear{{Tsukamoto}, {Takahashi}, {Machida}  \&
  {Inutsuka}}{{Tsukamoto} et~al.}{2015}]{TsukamotoEtal14}
{Tsukamoto} Y.,  {Takahashi} S.~Z.,  {Machida} M.~N.,   {Inutsuka} S.,  2015,
  \mn@doi [\mnras] {10.1093/mnras/stu2160}, \href
  {http://adsabs.harvard.edu/abs/2015MNRAS.446.1175T} {446, 1175}

\bibitem[\protect\citeauthoryear{{Vazan} \& {Helled}}{{Vazan} \&
  {Helled}}{2012}]{VazanHelled12}
{Vazan} A.,  {Helled} R.,  2012, \mn@doi [\apj] {10.1088/0004-637X/756/1/90},
  \href {http://adsabs.harvard.edu/abs/2012ApJ...756...90V} {756, 90}

\bibitem[\protect\citeauthoryear{{Vigan} et~al.,}{{Vigan}
  et~al.}{2012}]{ViganEtal12}
{Vigan} A.,  et~al., 2012, \mn@doi [\aap] {10.1051/0004-6361/201218991}, \href
  {http://adsabs.harvard.edu/abs/2012A%26A...544A...9V} {544, A9}

\bibitem[\protect\citeauthoryear{{Vorobyov} \& {Basu}}{{Vorobyov} \&
  {Basu}}{2005}]{VB05}
{Vorobyov} E.~I.,  {Basu} S.,  2005, \mn@doi [\apjl] {10.1086/498303}, \href
  {http://adsabs.harvard.edu/abs/2005ApJ...633L.137V} {633, L137}

\bibitem[\protect\citeauthoryear{{Vorobyov} \& {Basu}}{{Vorobyov} \&
  {Basu}}{2006}]{VB06}
{Vorobyov} E.~I.,  {Basu} S.,  2006, \mn@doi [\apj] {10.1086/507320}, \href
  {http://adsabs.harvard.edu/abs/2006ApJ...650..956V} {650, 956}

\bibitem[\protect\citeauthoryear{{Vorobyov} \& {Basu}}{{Vorobyov} \&
  {Basu}}{2010}]{VB10}
{Vorobyov} E.~I.,  {Basu} S.,  2010, \mn@doi [\apj]
  {10.1088/0004-637X/719/2/1896}, \href
  {http://adsabs.harvard.edu/abs/2010ApJ...719.1896V} {719, 1896}

\bibitem[\protect\citeauthoryear{{Young} \& {Clarke}}{{Young} \&
  {Clarke}}{2016}]{YoungClarke16}
{Young} M.~D.,  {Clarke} C.~J.,  2016, \mn@doi [\mnras]
  {10.1093/mnras/stv2378}, \href
  {http://adsabs.harvard.edu/abs/2016MNRAS.455.1438Y} {455, 1438}

\bibitem[\protect\citeauthoryear{{Zhu}, {Hartmann}  \& {Gammie}}{{Zhu}
  et~al.}{2009}]{ZhuEtal09}
{Zhu} Z.,  {Hartmann} L.,   {Gammie} C.,  2009, \mn@doi [\apj]
  {10.1088/0004-637X/694/2/1045}, \href
  {http://adsabs.harvard.edu/abs/2009ApJ...694.1045Z} {694, 1045}

\bibitem[\protect\citeauthoryear{{Zhu}, {Hartmann}, {Nelson}  \&
  {Gammie}}{{Zhu} et~al.}{2012}]{ZhuEtal12a}
{Zhu} Z.,  {Hartmann} L.,  {Nelson} R.~P.,   {Gammie} C.~F.,  2012, \mn@doi
  [\apj] {10.1088/0004-637X/746/1/110}, \href
  {http://adsabs.harvard.edu/abs/2012ApJ...746..110Z} {746, 110}

\makeatother
\end{thebibliography}



\bsp	
\label{lastpage}
\end{document}